\documentclass[useAMS,usenatbib]{mn2e}

\usepackage{amsmath}
\usepackage{amssymb}
\usepackage{upgreek}
\usepackage{longtable,lscape,psfig}
\usepackage[mathscr]{eucal}
\usepackage[dvips]{graphicx}

\title
[The Massive Star Population of Cygnus~OB2]
{The Massive Star Population of Cygnus~OB2}
\author
[Wright et al.]
{Nicholas J. Wright,$^1$ Janet E. Drew$^1$ and Michael Mohr-Smith$^1$\\
$^{1}$Centre for Astrophysics Research, University of Hertfordshire, Hatfield, AL10 9AB, UK\\
}

\begin{document}
\maketitle

\begin{abstract}

We have compiled a significantly updated and comprehensive census of massive stars in the nearby Cygnus OB2 association by gathering and homogenising data from across the literature. The census contains 169 primary OB stars, including 52 O-type stars and 3 Wolf-Rayet stars. Spectral types and photometry are used to place the stars in a Hertzprung-Russell diagram, which is compared to both non-rotating and rotating stellar evolution models, from which stellar masses and ages are calculated. The star formation history and mass function of the association are assessed, and both are found to be heavily influenced by the evolution of the most massive stars to their end states. We find that the mass function of the most massive stars is consistent with a `universal' power-law slope of $\Gamma = 1.3$. The age distribution inferred from stellar evolutionary models with rotation and the mass function suggest the majority of star formation occurred more or less continuously between 1 and 7~Myr ago, in agreement with studies of low- and intermediate mass stars in the association. We identify a nearby young pulsar and runaway O-type star that may have originated in Cyg~OB2 and suggest that the association has already seen its first supernova. Finally we use the census and mass function to calculate the total mass of the association of $16500^{+3800}_{-2800}$~M$_\odot$, at the low end, but consistent with, previous estimates of the total mass of Cyg~OB2. Despite this Cyg~OB2 is still one of the most massive groups of young stars known in our Galaxy making it a prime target for studies of star formation on the largest scales.

\end{abstract}


\section{Introduction}

Cygnus~OB2 is the nearest region to the Sun that can lay claim to being amongst the most massive clusters or associations in our galaxy \citep[e.g.,][]{knod00,hans03,wrig09a}. The region is known to harbour many tens of O-type stars and hundreds of OB stars \citep[e.g.,][]{mass91,come02,kimi07}, though estimates of the total mass of the association from both high- and low-mass stars range over an order of magnitude from $(2 - 10) \times 10^4$~M$_\odot$ \citep{knod00,hans03,drew08,wrig10a}. The association is itself the most obvious example of recent star formation in the massive Cygnus~X complex, a region of ongoing star formation \citep{reip08} that spans $\sim$200~pc at a distance of $1.4 \pm 0.08$~kpc \citep{rygl12}. The size and proximity of Cyg~OB2 therefore make it a template for studies of distant, massive star clusters and OB associations in the Milky Way and in nearby galaxies. 

\begin{figure*}
\begin{center}
\includegraphics[width=500pt]{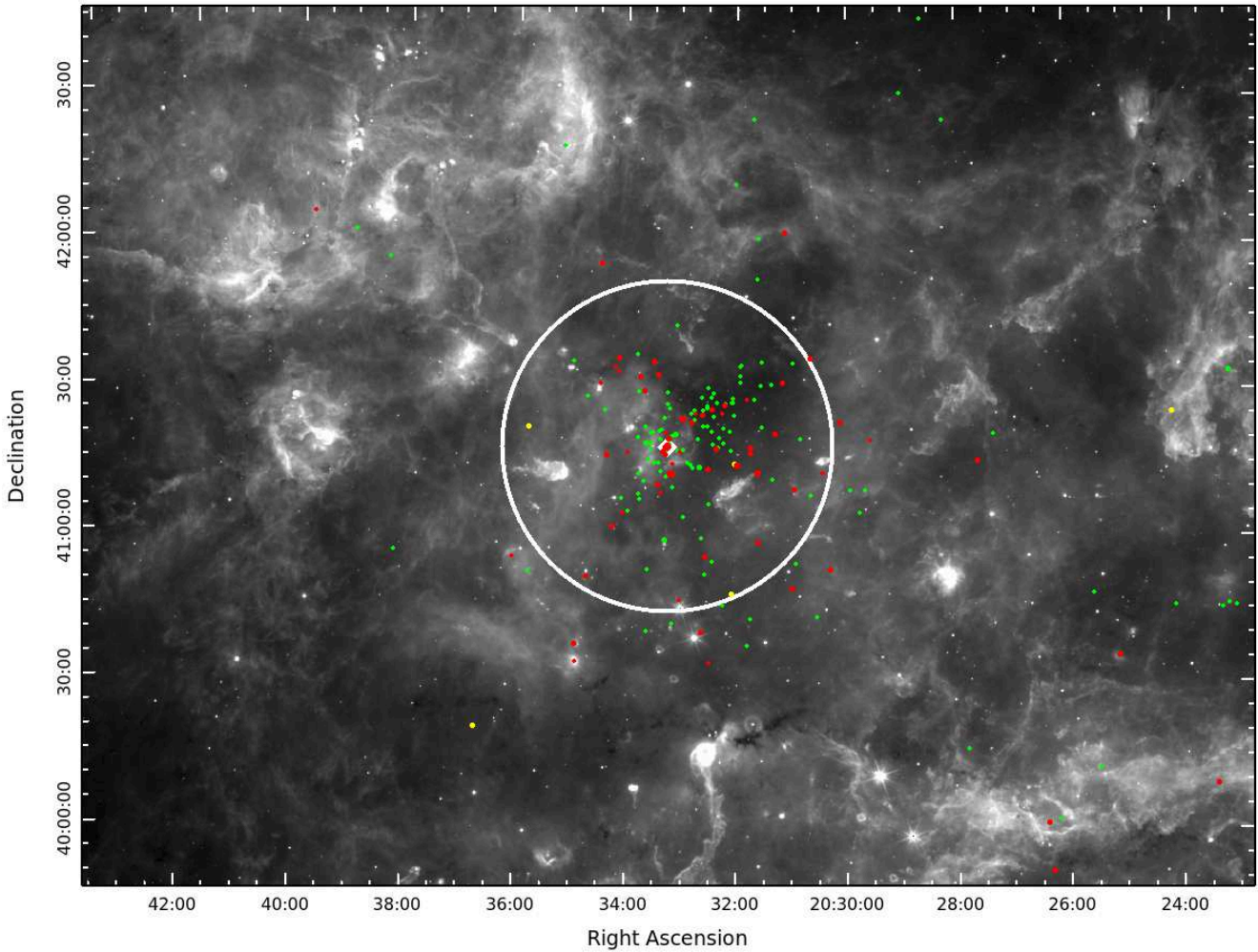}
\caption{Mid-infrared 12~$\mu$m image of the Cygnus~X giant molecular cloud compiled from Wide-Field Infrared Survey \citep{wrig10c} data and spanning $4 \times 3$ degrees ($100 \times 75$~pc at the distance of Cyg~OB2). Known O, B and Wolf Rayet stars are shown as red, green and yellow dots respectively (compiled from the literature as outlined in Section~\ref{compilation}. The white circle shows the 1 square degree region studied in this work, which is centred on the ``trapezium'' of O stars known as Cyg~OB2 \#8, itself marked with a white diamond, and as broadly co-spatial with the Cygnus~OB2 {\it Chandra} Legacy Survey \citep{wrig14c}.}
\label{map}
\end{center}
\end{figure*}

The high-mass stellar content of Cyg~OB2 has been well studied over the past half century since the association was first identified by \citet{munc53} and a number of mid- to late-O stars within it were classified by \citet{john54}. Further members were identified and classified by \citet{morg54}, \citet{schu56}, and \citet{schu58}. While many of these classifications were revised over the following decades \citep[e.g.,][]{walb73}, it wasn't until the work of \citet{mass91} that a more extensive spectral survey of the region was conducted, resulting in the identification of 42 O-type stars and 26 B-type stars. A decade later the number of known OB stars increased again with the works of \citet{come02} and \citet{hans03} following the photometric study of \citet{knod00}. Since then the number of very massive O-type stars has not increased significantly, with the more recent and deeper surveys uncovering predominantly B-type stars \citep[e.g.,][]{kimi07,kobu12,come12}. This apparent dearth of new O-type stars discovered in Cyg~OB2, despite deeper observations, particularly in the near-IR, and through spectroscopic observing campaigns, may suggest that our census of massive O-type stars is nearing completion.

The age of Cyg~OB2 has proved to be a complex subject, due in no small part to the large spatial size of the association and the lack of any clearly recognisable limit to its extent. \citet{hans03} compiled a sample of 85 OB stars in Cyg~OB2, from which the presence of a number of mid O-type dwarf stars and high luminosity blue supergiants led her to conclude that the age of Cyg~OB2 could not be much older than a few million years, and settled on an age of 2~Myrs, with a spread of 1~Myr. The evolved giant and supergiant stars present were argued by \citet{hans03} to be contaminating non-members, brought into the association as the search for cluster members was extended over a larger area \citep[e.g.,][]{come02}. However, studies of A-type stars in Cyg~OB2 have revealed a population of 5--7~Myr old stars toward and to the south of the association \citep{drew08}, while X-ray studies of low-mass stars in the association suggest an age of 3--5~Myrs \citep{wrig10a}. If the OB stars in Cyg~OB2 are younger than the low-mass stars it could suggest that the massive stars formed after the low-mass stars, a feature observed in some other young clusters \citep[e.g.,][]{povi10} that has interesting implications for theories of massive star formation and feedback.

The initial mass function (IMF) of high-mass stars provides a powerful diagnostic of the star formation process and is also therefore an area of considerable interest, particularly as to whether the IMF varies between OB associations and star clusters \citep[e.g.,][]{bast10}, particularly given recent suggestions that Cyg~OB2 has always been a low-density association and did not evolve from a star cluster. \citet{mass91} found the high-mass IMF in Cyg~OB2 to be considerably flatter than found for other regions in the galaxy, though \citet{kimi07} found it to be much steeper than the canonical Salpeter IMF \citep{salp55} from an expanded sample, while \citet{wrig10a} found the lower-mass IMF to be in agreement with the Salpeter value.

Work over the past decade has assembled the data needed to form a much fuller view of the OB population in Cyg~OB2, thanks to a combination of new spectroscopic surveys of both primary OB stars and their binary companions \citep[e.g.,][]{kimi07,kobu12}, improvements in the stellar effective temperature scale that consider line-blanketing and non-LTE (local thermodynamic equilibrium) effects \citep{mart05}, and the availability of revised stellar evolution models for massive stars that take into account the important effects of stellar mass-loss and rotation \citep{ekst12}. \citet{weid10} found that masses derived from rotating stellar evolutionary models were in much better agreement with dynamical masses for stars in binary systems than were masses derived from non-rotating stellar models, which may go some way towards addressing previous reports of a non-standard initial mass function (IMF) in Cyg~OB2. Finally, recent high-spatial resolution optical and near-IR Galactic Plane surveys \citep{drew05,luca08,groo09} provide the necessary photometry to resolve close and blended stars in the commonly used catalogs of \citet{mass91} and the 2 Micron All Sky Survey \citep[2MASS,][]{skru06} - both of which suffer from some degree of blending that is occasionally ignored by users of the data.

Our objective for this work is to bring together all available information on the massive star population in Cyg~OB2, gathering spectroscopy from across the literature and selecting the best available photometry for these sources. Our hope is that this will allow a quantitative analysis of the star formation history, initial mass function, and total mass of Cyg~OB2. In Section~2 we outline the compilation of spectroscopic and photometric data from the literature for our census of 169 OB stars, almost doubling the sample of 85 OB stars studied by \citet{hans03}. In Section~3 we use this data to calibrate the extinction towards Cyg~OB2 and verify the extinction law for this sightline that is then used to derive individual extinctions towards the massive stars. In Section~4 we compile a Hertzsprung-Russell (HR) diagram for the population of massive stars, from which individual stellar masses and ages are derived. In Section~5 we study the age distribution of these stars to infer the star formation history of the association and in Section~6 we study their mass function. In Section~7 we use these results to estimate the total stellar mass of the association, and finally in Section~8 we summarise our conclusions.

\section{Compilation of the sample}

In this section we compile the available photometry and spectroscopy from the literature. Figure~\ref{map} shows the distribution of known OB stars in Cyg~OB2 and the wider Cygnus~X region centered on the densest part of the association. One of the difficulties with compiling a census of Cyg~OB2 is that as an association and not a centrally-condensed cluster it is not immediately obvious where to set the limits of such a study. We focus on a 1 square degree circular area centred on Cyg~OB2 \#8 - the trapezium of O stars at RA 20:33:16, Dec +41:18:45 long regarded as being at the heart of the association \citep[e.g.,][]{hans03,vink08}. This is equivalent to a radius of $\sim$14~pc at the distance of Cyg~OB2. This choice has a number of advantages, most notably that it includes the clear overdensity that constitutes Cyg~OB2, but does not extend too far into the `field' population of OB stars that exists across Cygnus~X \citep[e.g.,][]{come08}.

\subsection{Spectroscopic data}
\label{compilation}

The census was compiled by searching the literature for all spectroscopically classified stars of spectral type B5 or earlier, a range of spectral types that can provide a solid diagnosis of the age, mass function and spatial extent of the association. This resulted in 167 stars (22 with known binary companions) from 18 different publications. To limit any potential discrepancies arising from using stars classified by different authors wherever possible we use classifications made by comparison with the stellar atlas of \citet{walb90}. Fortunately this includes the majority of known OB stars in Cyg~OB2. When a star is classified in this way in multiple papers the practice here is to adopt the classification made at the highest spectral resolution.

The majority of stars come from the studies of \citet[][50 stars, or 26\% of both primary and secondary stars]{mass91} and \citet[][78 stars or 41\%]{kimi07}, which show excellent agreement with each other when their classifications overlap. A further 27 stars (14\%) were taken from a series of papers by the latter team of authors \citep{kimi08,kimi09,kimi12,kobu12}. The remaining 37 spectral classifications (19\%) used in this work come from 12 different studies with between 1 and 7 classifications each, including many studies devoted to a single particularly interesting star. Classifications from these focussed studies were favoured where possible as they often uncovered previously unknown binary companions that changed the classification of the primary star.

Spectral classifications are presented in Table~\ref{obstars} with references, known binary companions, and source numbers under the systems initiated by \citet[][the Schulte system, e.g., Cyg~OB2 \#5]{john54}, \citet[][e.g., MT426]{mass91}, and \citet[][e.g., A26]{come02}. We note that the star listed as `A11' by \citet{come02} is actually MT267 from \citet{mass91}, and is listed as the latter in this work. By comparing spectral classifications of the same star performed by different authors we estimate the uncertainty in the spectral types presented here to be approximately half a subtype for O and early B-type (B0-B1) stars and one subtype for later B-type stars. The standard deviation in luminosity class is approximately half a luminosity class for O-type stars, one luminosity class for early B-type stars (B0-B2), and may be higher for later B-type stars, but there is insufficient data for a meaningful comparison. These uncertainties are used in all following calculations to derive uncertainties on other parameters.

\subsection{Model atmosphere fits}

In this work physical parameters for stars will mainly be based on spectral types from the literature, using the effective temperature scales of \citet{mart05}, \citet{trun07}, and \citet{hump84}. However, effective temperatures derived from model atmosphere fits should be more reliable, both because they are not quantised in spectral subtypes (or half subtypes), but also because they take into account other parameters such as stellar mass-loss and chemical abundances.

We searched the literature and identified 11 stars with suitable model atmosphere fits from the works of \citet{herr02} and \citet{negu08}. We did not include the model atmosphere fits from \citet{herr99} since these used mass-loss rates and line blanketing analyses that have since been superseded (Herrero, private communication). We have also excluded stars that have since been found to be binaries with similar luminosities, as these will have blended spectra that will affect the model atmosphere fits. 

\begin{table*}
\begin{center}
\caption{Model atmosphere fits used in this work} 
\label{modelfits}
\begin{tabular}{lcccccccccr}
\hline
Name		& \multicolumn{4}{c}{Spectral types}	&& \multicolumn{5}{c}{Model atmosphere fit} \\
\cline{2-5} \cline{7-11}
	 		& Type	& Ref& $T_{eff} (K)$ & log~$g$	&& $T_{eff}$ (K) & $\sigma_{T_{eff}}$ (K) & log~$g$	& $\sigma_{\mathrm{log} \, g}$	& Ref	 \\
\hline
\#2			& B1I	& MT91	& 24,300		& -		&& 28,000	& 1000		& 3.21	& 0.1				& H02\\
\#4			& O7III	& K07	& 36,077		& 3.61	&& 35,500	& 1000		& 3.52	& 0.1				& H02\\
\#8C			& O5III	& K07	& 40,307		& 3.69	&& 41,000	& 1500		& 3.81	& 0.1				& H02\\
\#10			& O9I	& K07	& 31,368		& 3.23	&& 29,000	& 1000		& 3.11	& 0.1				& H02\\
\#11			& O5If+B0V & K12a	& 38,612		& 3.57	&& 37,000	& 1500		& 3.61	& $^{+0.15}_{-0.10}$	& H02\\
MT267 (A11)	& O7.5III	& K12a	& 35,019		& 3.59	&& 36,000	& 1500		& 3.6		& 0.2				& N08\\
A15			& O7I	& N08	& 34,990		& 3.40	&& 35,000	& 1000		& 3.2		& 0.2				& N08\\
A24			& O6.5III	& N08	& 37,134		& 3.63	&& 37,500	& 1500		& 3.6		& 0.2				& N08\\
A26			& O9.5V	& N08	& 31,884		& 3.92	&& 35,000	& 1000		& 3.9		& 0.2				& N08\\
A33			& B0.2V	& N08	& 29,390		& -		&& 31,000	& 1500		& 4.0		& 0.2				& N08\\
A38			& O8V	& N08	& 34,877		& 3.92	&& 36,000	& 1500		& 4.0		& 0.2				& N08\\
\hline
\end{tabular} 
\end{center}
\flushleft
The 11 stars with model atmosphere fits used in this work, with effective temperatures and gravities taken from the fits and calculated from the spectral types listed. Effective temperatures and gravities calculated as a function of spectral type using the data from \citet{mart05} and \citet{trun07} where possible. Uncertainties on the spectral type parameters are estimated from the spectral type and luminosity class uncertainties estimated in Section~\ref{compilation} and are approximately $\pm$1000~K (approximately $\pm$3000~K for the early B-type stars) and $\pm$0.1~dex. Model atmosphere fits taken from \citet[][H02]{herr02} and \citet[][N08]{negu08}, and spectral types from \citet[][MT91]{mass91}, \citet[][K07]{kimi07}, and \citet[][K12a]{kobu12}.
\end{table*}

The 11 stars with model atmosphere fits are listed in Table~\ref{modelfits} with the fitted effective temperatures and gravities. Compared to the effective temperatures and gravities determined from the spectral types (also listed in Table~\ref{modelfits}) there is good agreement within the uncertainties. Given that both methods employ the same effective temperature scale, this is perhaps not surprising.

For the remainder of this work for the 11 stars listed in Table~\ref{modelfits} we used the effective temperatures derived from the model atmosphere fits and then calculated intrinsic colours and bolometric corrections (BC) for these stars as a function of effective temperature (instead of spectral type), as described below.

\subsection{Photometric data}

Photometric data in the optical and near-IR was gathered from the literature to complement the spectroscopy. In the optical the primary source of photometry was the $UBV$ data from \citet{mass91}, which is commonly used for studies of Cyg~OB2. However, some of these photometric measurements are tagged by \citet{mass91} as `blended', including 11 stars in our sample. We therefore supplemented the $UBV$ data with $U g^\prime r^\prime$ photometry from the Ultra-violet Excess Survey \citep[UVEX,][]{groo09} and converted this onto the $UBV$ system (see Appendix~\ref{s-blended}). For the 11 `blended' sources, 7 had sufficient photometry to calculate full $UBV$ photometry while 4 had only $U$ and $g^\prime$ UVEX photometry ($r^\prime$ was saturated) and so only $UB$ photometry was calculated. In addition 12 sources in our sample were not from the original \citet{mass91} sample \citep[predominantly from the near-IR sample gathered by][]{come02} and we therefore sought UVEX photometry to homogenise our sample. Of these 12 sources, 9 had sufficient photometry to derive $U$ and $B$ photometry, while the remaining 3 were either saturated in the UVEX images or the sources were not observed. For all these sources the photometric uncertainties used were propagated from the UVEX photometric uncertainties and the dispersions on the transformations used.

Near-IR photometry was taken from 2MASS for the majority of stars. In view of the low spatial resolution of the 2MASS observations that could lead to blending in some cases we sought replacement photometry from the  UKIDSS-GPS \citep[United Kingdom Infrared Deep Sky Survey Galactic Plane Survey,][]{luca08}, but were only able to obtain unsaturated photometry for 3 sources. For these sources we replaced their 2MASS photometry by UKIDSS photometry using the transformations in \citet{luca08}.

The final photometric data used in this study is presented in Table~\ref{obstars} on the $UBV$ and 2MASS $JHK_s$ photometric systems. The vast majority (93\% or 155 out of 167) of the primary stars have optical photometry in at least two of the $UBV$ bands, and all of the targets have near-IR photometry in $J$, $H$, and $K_s$.

\subsection{Binary stars and their influence on stellar parameters}

Unresolved binary systems can lead to the miscalculation of the properties of the primary star. The contribution of light from a binary companion, which in involved systems is cooler than the primary, will make the primary appear cooler and more luminous, which could potentially lead to the star being diagnosed as more evolved and higher-mass. The binary population of massive stars in Cyg~OB2 has recently been well studied \citep[e.g.,][]{kimi07,kimi09,kimi12,caba14} and 22 of our 167 primary stars have secondaries with known spectral types, although it is likely that there still remain a number of undiscovered binary companions in this sample.

For objects with known companions it is possible to correct for the light of the secondary when using the observed photometry as long as the spectral type of both stars is known. This is relatively simple when the luminosity ratio of the binary system is known in all the relevant bands, or if the bolometric luminosity ratio is known and the stars are of similar spectral type \citep[e.g., Cyg~OB2 \#5 is an O7I+O6I binary with a bolometric luminosity ratio of 3.1,][]{lind09}. For other binary systems this is trickier and we must make an assumption about the relative in-band fluxes of the two stars, for which the simplest approach and that adopted here is to assume that the ratio of fluxes is equal to the ratio of intrinsic fluxes according to spectral type from \citet{mart05} for O-type stars and \citet{cont08} for B-type stars.

Doing this for our sample we find that the typical (rms) adjustment to the colour of the primary star is very small, 0.006~mag, having a negligible effect on the extinctions. This is because the strongest effect occurs when two stars have significantly different intrinsic colours, but in such systems the stars usually have very different luminosities and therefore the influence of the secondary is small. The rms increase in the observed magnitude is more significant at 0.5~mag, equivalent to a 37\% reduction in the bolometric luminosity. This luminosity reduction (magnitude increase) varies from 8\% (0.09) for Cyg~OB2 \#11 (O5If+B0V) to 50\% (0.75) for very similar pairs of stars such as Cyg~OB2 \#73 (O8III+O8III) and B17, an O7I+O9I binary.

Cyg~OB2 does contain one known triple system, MT429, which \citet{kimi12} find to be a B0V+B3V+B6V system. For simplicity we consider only the two brightest components of the system because the B6V star is too faint to affect the colour or luminosity of the primary star. We also note that while Cyg~OB2 \#22, an O3If+O6V binary \citep{walb02}, has been resolved in high angular resolution images with a separation of $\sim$1.5$^{\prime\prime}$ between the two components \citep{maso09,maiz10}, these observations do not provide sufficient photometry to place both components on the HR diagram separately, therefore we only consider the primary star in this work (after applying the corrections for the secondary noted above).

\subsection{Completeness}

It is difficult to estimate the completeness of a sample that is compiled from so many different sources of spectroscopy, each of which is based on a different method of photometric selection, area of study, and spectroscopic completeness. However, one possible indicator of the level of completeness of our sample is the fact that the most recent spectroscopic studies \citep[e.g.,][]{kimi07,kobu12} have discovered only B-type stars and not new O-type stars, possibly suggesting that the majority of O-type stars in Cyg~OB2 have already been detected. The fact that the most recent studies have found stars with similar extinctions to previous studies \citep{come12} also suggests that a previously undiscovered population of highly obscured ($A_V > 10$~mag) O-type stars does not exist. Based upon this we will make the assumption that our census of massive stars is complete for O-type stars in Cyg~OB2, which is broadly equivalent to being complete down to 15~M$_\odot$ for zero-age main-sequence stars, or 20~M$_\odot$ if one takes into account some further evolution. In a future paper we will attempt to test this assumption about our completeness using new spectroscopic observations of Cyg~OB2.

\section{The extinction towards Cygnus~OB2}
\label{s-extinctions}

In this section the spectral types and photometry are used to re-derive the form of the extinction law towards Cyg~OB2 and then calculate individual extinctions for all the stars in our sample. To achieve this we use the unreddened colours from \citet{mart06} for O-type stars (the ``observational'' $T_{eff}$ scale) and from \citet{fitz70} and \citet{koor83} for B-type stars, adjusted for B0-B1 stars to provide a smooth transition with the O star colours.

\subsection{The extinction law towards Cygnus~OB2}

\begin{table}
\begin{center}
\caption{Stellar sample used to determine the extinction law towards Cyg OB2.}
\label{reddeningsample}
\begin{tabular}{llcc}
\hline  
Spectral type & Number & $R_V$ & $A_V$ \\ 
\hline 
O6V & 005 & $ 2.95 \pm 0.04 $ & $ 6.00 \pm 0.04 $ \\ 
O9V & \#14 & $ 2.88 \pm 0.04 $ & $ 4.49 \pm 0.03 $ \\ 
O8V & \#6 & $ 2.89 \pm 0.04 $ & $ 4.48 \pm 0.03 $ \\ 
O8.5V & \#17 & $ 2.93 \pm 0.04 $ & $ 4.88 \pm 0.03 $ \\ 
O8V & 376 & $ 2.89 \pm 0.04 $ & $ 4.85 \pm 0.03 $ \\ 
O8V & 390 & $ 2.83 \pm 0.04 $ & $ 6.49 \pm 0.04 $ \\ 
O8V & 455 & $ 2.85 \pm 0.04 $ & $ 6.01 \pm 0.04 $ \\ 
O9.5V & \#23 & $ 2.90 \pm 0.04 $ & $ 5.06 \pm 0.04 $ \\ 
O8.5V & \#8D & $ 2.86 \pm 0.04 $ & $ 5.07 \pm 0.04 $ \\ 
O7.5V & \#24 & $ 2.94 \pm 0.04 $ & $ 5.59 \pm 0.03 $ \\ 
O8V & 485 & $ 2.87 \pm 0.04 $ & $ 5.25 \pm 0.03 $ \\ 
O8.5V & 507 & $ 2.89 \pm 0.04 $ & $ 5.34 \pm 0.04 $ \\ 
O7.5V & 534 & $ 2.88 \pm 0.04 $ & $ 6.26 \pm 0.04 $ \\ 
O8V & \#74 & $ 2.89 \pm 0.03 $ & $ 6.36 \pm 0.04 $ \\ 
O7V & 611 & $ 2.97 \pm 0.04 $ & $ 5.51 \pm 0.04 $ \\ 
O9V & \#41 & $ 2.84 \pm 0.04 $ & $ 6.04 \pm 0.04 $ \\ 
O9V & \#75 & $ 3.08 \pm 0.04 $ & $ 5.48 \pm 0.04 $ \\ 
O7V & \#29 & $ 3.02 \pm 0.04 $ & $ 5.35 \pm 0.03 $ \\ 
\hline
\end{tabular} 
\newline
All stars classified by \citet{mass91} with source numbers from that work, except those noted with \#, which are labelled on the Schulte system. Note that while MT005 and MT611 have been observed to have binary companions \citep{caba14}, the companions are significantly fainter than the primary stars ($\delta V = 2.8$ and $\delta K = 4.9$, respectively) and are therefore unlikely to affect the SEDs of the primary stars.
\end{center}
\end{table}

To determine the form of the extinction law towards Cyg~OB2 we selected a high-confidence subset of 18 stars from within our sample to which we perform spectral energy distribution (SED) fitting. We selected only single O-type main-sequence stars classified by \citet{mass91} and with no evidence of having a similar-luminosity binary companion. By using only O-type stars we reduce the uncertainty due to the effective temperature and the binarity of the stars, both of which increase towards later spectral types. In addition we required all stars to have been detected in, and have low errors ($< 0.05$ in $UBV$ and $< 0.02$ in $JHK_s$) in, all six photometric bands. These criteria resulted in the sample of 18 stars in Table~\ref{reddeningsample}.

Reddened stellar spectra were calculated using CMFGEN non-LTE model spectra \citep{hill98c}, as used by \citet{mart05}, reddened using the extinction curves presented by \citet{fitz07} as a function of $R_V$ and $A_V$. Reddening coefficients were determined using Kitt Peak $UBV$ filter profiles \citep[appropriate for the observations of][]{mass91} and 2MASS $JHK_s$ filter profiles \citep{skru06}. Combining these with the intrinsic colours for O-type dwarf stars presented by \citet{mart06} results in reddened photometry as a function of spectral type, $R_V$, and $A_V$.

\begin{figure}
\begin{center}
\includegraphics[height=240pt, angle=270]{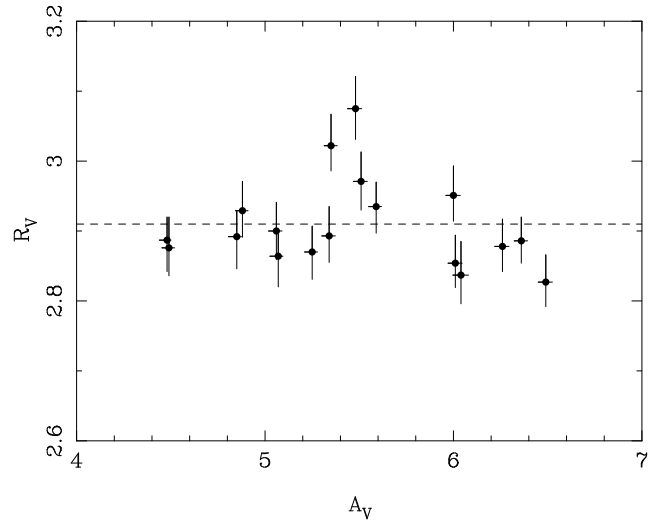}
\caption{$R_V$ versus $A_V$ for our high-confidence sample of 18 single O-type dwarfs derived using $\chi^2$ fits between the observed photometry and reddened model spectra using the \citet{fitz07} reddening curves. 1$\sigma$ error bars are shown derived from the $\chi^2$ fits. The weighted mean $R_V = 2.91 \pm 0.06$ is shown as a dashed line.}
\label{rv_results}
\end{center}
\end{figure}

$\chi^2$ fits were performed between the observed photometry and the reddened model photometry allowing $R_V$ to vary in steps of 0.01 in the range 2.5--3.5 and with $A_V$ varying between 4.0--7.0~mag in steps of 0.01~mag. The distance modulus was left as a free parameter to account for variations in the intrinsic distance and luminosity of stars. The best fits are shown in Figure~\ref{rv_results} and listed in Table~\ref{reddeningsample}.

The weighted mean extinction law from our $\chi^2$ fits is $R_V = 2.91 \pm 0.06$ (taken as such for the remainder of this work), in good agreement with previous studies based on optical colours \citep[e.g.,][measured $R_V = 3.0$]{hans03}. However this value is in disagreement with the findings of \citet{come12} who use a combination of optical and near-IR photometry to derive $R_V = 4.0$ using a \citet{card89} extinction law for OB stars over a wider area across the Cygnus~X molecular cloud. This disagreement is likely explained by there being two different extinction laws at work along different sight lines: a `standard' $R_V \sim 3$ extinction law towards Cyg~OB2 \citep[due to obscuration from the Cygnus Rift, which lies predominantly in the foreground,][]{guar12}, and an `anomalous' $R_V \sim 4$ extinction law away from Cyg~OB2, possible due to a change in the grain size distribution favouring larger dust grains within the Cygnus~X molecular cloud (which has already been dispersed from the vicinity of Cyg~OB2, see Figure~\ref{map}). A similar picture of two extinction laws has been observed by \citet{povi11} towards the Carina molecular cloud.

For reference this new extinction law is presented in Table~\ref{table_extinction} for extinctions between $A_V = 4.0$--7.0 (the typical range observed in Cyg~OB2, see Section~\ref{s-extstars}) and compared to that found previously by \citet{hans03} from comparison of $UBVJHK_s$ colour excesses compared to the \citet{leje01} intrinsic colours. The extinction law is relatively unchanged in the optical, but has changed significantly in the near-IR, due mostly to shifts in the near-IR intrinsic colours of O-type stars between the works of \citet{leje01} and \citet{mart06}.

\begin{table}
\begin{center}
\caption{Extinction law towards Cyg~OB2} 
\label{table_extinction}
\begin{tabular}{lcccc}
\hline 
Band ($\lambda$) & \multicolumn{4}{c}{$A_\lambda / A_V$}\\
\cline{2-5}
 & \multicolumn{3}{c}{This work} & Hanson \\
\cline{2-4}
 & $A_V = 4.0$ & $A_V = 5.5$ & $A_V = 7.0$ \\
\hline
$U$ (3372 \AA)			& 1.593	& 1.599 	& 1.605	& 1.600 \\
$B$ (4404 \AA)			& 1.321	& 1.318 	& 1.316	& 1.333 \\
$V$ (5428 \AA)			& 1.000	& 1.000 	& 1.000	& 1.000 \\
$J$ (1.27 $\mu$m) 		& 0.233	& 0.234 	& 0.235	& 0.282 \\
$H$ (1.67 $\mu$m) 		& 0.138	& 0.139 	& 0.140	& 0.175 \\
$K_s$ (2.16 $\mu$m) 	& 0.0840	& 0.0845 	& 0.0850	& 0.125 \\
\hline
\end{tabular} 
\newline
\end{center}
\flushleft
New extinction law for Cyg~OB2 calculated using a \citet{fitz07} $R_V = 2.91$ extinction curve and various extinctions. The previous $R_V = 3.0$ extinction law presented by \citet{hans03} is given for reference.
\end{table}

\subsection{The extinction of Cygnus~OB2 members}
\label{s-extstars}

Using the \citet{fitz07} $R_V = 2.91$ extinction law derived above we calculated individual extinctions for the 164 primary O and B-type stars in our sample (not including the WR stars) from the comparison between observed and intrinsic colours. We favoured the $B-K_s$ colour for this due to its long baseline and the availability of the relevant photometry for the vast majority (160 or 98\%) of our sources. For 1 star $B$-band photometry was unavailable so we used the $V-K_s$ colour and for 3 stars only near-IR photometry was available so we used the $J-K_s$ colour. Uncertainties in the derived extinction were calculated using a Monte Carlo (MC) simulation taking into account the uncertainties in the photometry, spectral classification, the intrinsic colours (assumed to be half the difference between the colours of neighbouring subtypes), and the reddening law of $R_V = 2.91 \pm 0.06$. The typical uncertainty in the extinction is $\sim$0.3~mag in the $V$-band. Extinctions and uncertainties are presented in Table~\ref{obstars}.

\begin{figure}
\begin{center}
\includegraphics[height=240pt, angle=270]{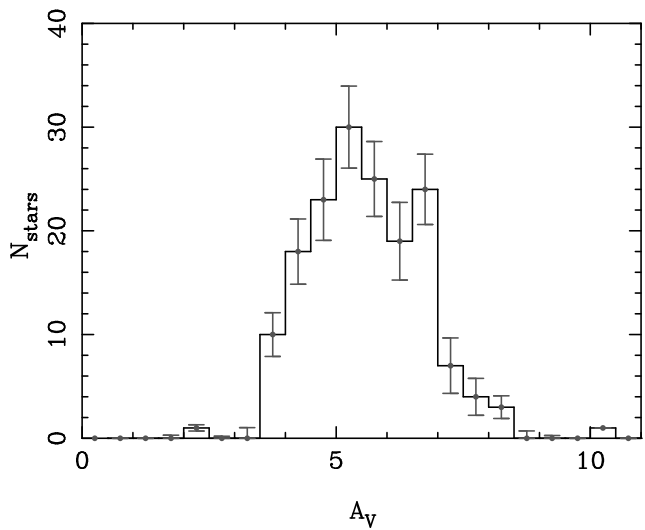}
\caption{Histogram of the extinctions calculated for all 164 massive OB stars in this work with 1$\sigma$ error bars derived from MC simulations.}
\label{extinction_histogram}
\end{center}
\end{figure}

Figure~\ref{extinction_histogram} shows the extinction distribution, which varies from $A_V = 2.2$~mag for MT140 (an O9.5I star on the western side of the association) to $A_V = 10.2$~mag for the central blue hypergiant Cyg~OB2~\#12 \citep[a B3.5Ie star, and one of the most luminous known stars in the Milky Way,][]{morg54,clar12}, though the majority lie in the range $A_V =$~4--7~mags with a median of $A_V  = 5.4$~mag and 25\% and 75\% quartiles of 4.7 and 6.5~mags. Error bars on the extinction distribution histogram were derived from the individual uncertainties using a MC simulation. The large spread in extinction compared to the typical uncertainty in $A_V$ suggests that the spread is real and is either due to foreground extinction from the Cygnus Rift or intra-association extinction within Cyg~OB2.

\begin{figure}
\begin{center}
\includegraphics[width=240pt]{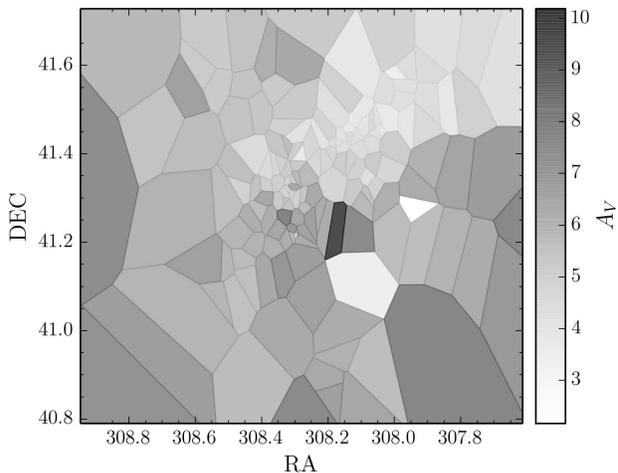}
\caption{Voronoi tessellated extinction map (one star per Voronoi cell) for for all 164 massive stars in Cyg~OB2 with the extinction indicated by the greyscale colour. The circular border of our study is approximately at the border of this image.}
\label{extinction_voronoi}
\end{center}
\end{figure}

Figure~\ref{extinction_voronoi} shows a spatial map of the extinction across Cyg~OB2 in the form of a Voronoi tessellation. The extinction distribution varies smoothly across the association, from $A_V \sim 4$--5~mag in the north-west \citep[where a known `reddening hole' exists,][]{redd67} to $A_V \sim 6$--7 to the south of the association and on the outskirts.

These results are in good agreement with previous studies of both the high- \citep{hans03} and low-mass populations \citep{wrig10a}, albeit with a smaller spread than the latter, likely due to the higher precision afforded by having known spectral types and a calibrated extinction law. The extinction is larger than that derived by \citet{guar12} from their $riz$ study by $\sim$1~mag (they derive a median extinction of $A_V = 4.3$~mag), though we observe the same spatial variation in extinction that they do. The anomalously low extinction derived by \citet{guar12}, particularly compared to many other studies, may be due to the difficulties deriving extinction from isochrone fits in the $riz$ filter system that requires a complex transformation from $BVI$ colours.

The high extinction for Cyg~OB2~\#12, $>2$~mags larger than any of the stars surrounding it, is particularly apparent. Some of this extinction may be due to circumstellar material given the advanced evolutionary stage of this object and that material is known to surround other very massive stars \citep[e.g.,][]{wrig14a}, however \citet{clar12} could find no evidence for a near- to mid-IR excess in the spectral energy distribution of this object, arguing against a significant amount of warm circumstellar dust. It is also possible that the SED we have assumed is wrong, given the unusual character of the star.

\section{Hertzprung-Russell Diagram for Cyg~OB2}

\begin{figure*}
\begin{center}
\includegraphics[height=495pt, angle=270]{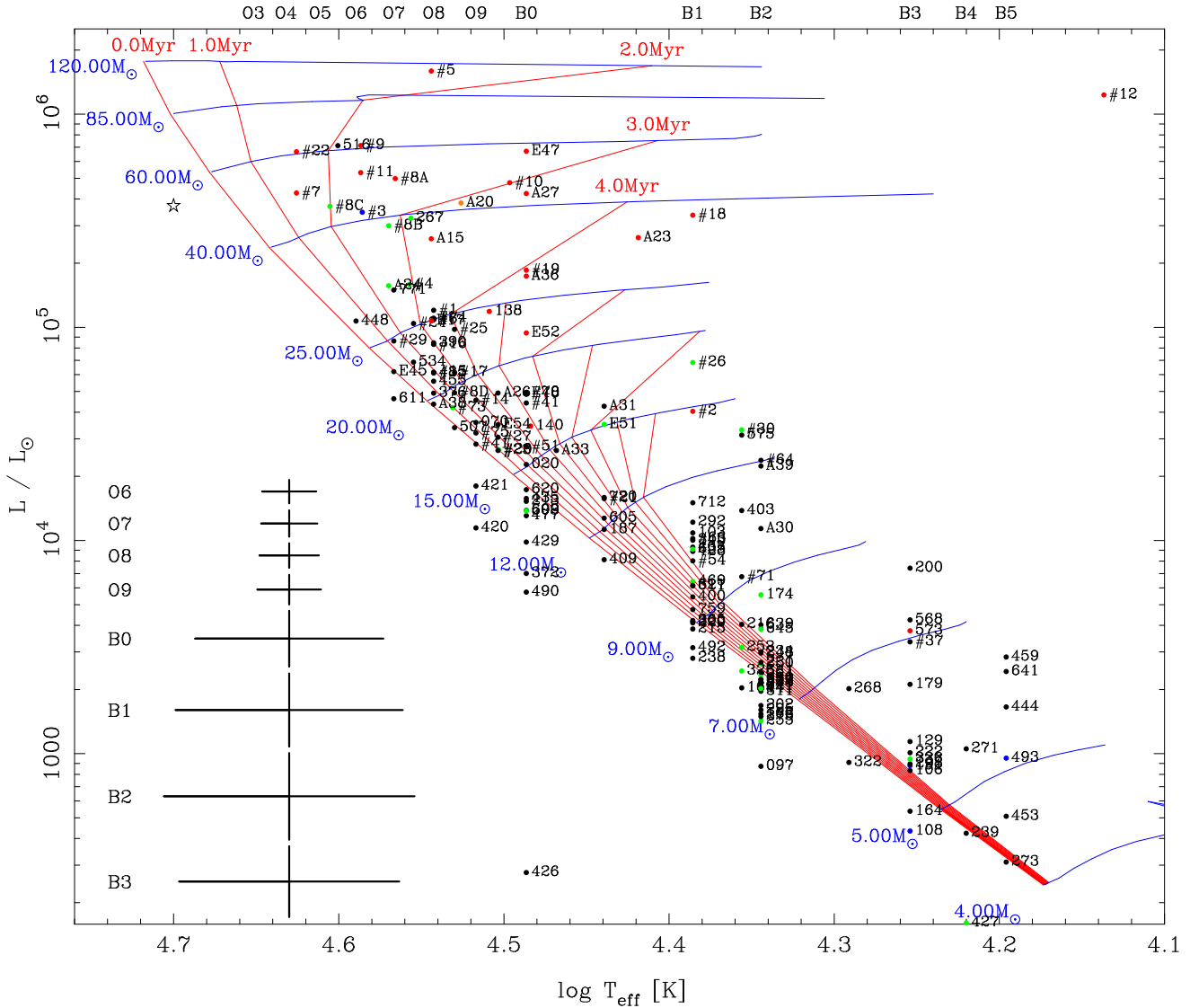}
\caption{HR diagram for Cyg~OB2 showing all the stars compiled in this work colour-coded by their luminosity class (dwarfs are black, subgiants are blue, giants are green, bright giants are orange and supergiants are red, while WR stars are shown with a star symbol) and with source numbers noted according to the scheme adopted in this paper. The figure shows isochrones and evolutionary tracks from \citet{leje01} with ages and masses labelled. The effective temperature scales used for main-sequence O- and B-type stars are shown along the top of the diagram and mean 1$\sigma$ errors bars for each subtype are shown to the bottom left.}
\label{HRD_lejeune}
\end{center}
\end{figure*}

In this section we place the stars on the HR diagram and compare their positions with both rotating and non-rotating stellar evolutionary tracks to derive stellar masses and ages. Cyg~OB2 also includes 4 evolved massive stars (such as Wolf Rayet stars) whose positions on the HR diagram and resulting stellar properties cannot be derived in the same way as the less evolved OB stars. We have gathered the properties of these stars from the literature, adjusting for a distance of 1.45~kpc if necessary, and derive additional physical properties (such as upper or lower limits on stellar ages of masses) by comparison with the evolutionary models used in this work or from our current, but far from complete, understanding of the evolution of massive stars. These stars and their properties are included in Table~\ref{obstars} and included in all later analysis where possible.

\subsection{Calculation of $L_{bol}$ and $T_{eff}$}

Bolometric luminosities were calculated using the observed $K_s$ band magnitudes because of the high extinction towards Cyg~OB2 and the typical uncertainty of 0.3~mag in $A_V$ resulting from this. The considerably lower uncertainty in $A_K$ compared to $A_V$ is preferable despite the higher uncertainty in $BC_K$ over $BC_V$ \citep[due to the strong sensitivity of the spectral energy distribution to wind parameters at longer wavelengths,][]{mart06}. BC were taken from \citet{mart06} for O-type stars, \citet{crow06} for B-type supergiants, and \citet{hump84} for less luminous B-type stars, although there is very little variation with luminosity class. Effective temperatures were determined as a function of spectral type using the tabulations of \citet[][``observed'' scale]{mart05} for O-type stars, \citet{trun07} for early (B0-3) B-type stars and \citet{hump84} for later B-type stars, all of which show a good agreement between each other.

The accepted distance to Cyg~OB2 has been reduced over the last few decades due to improved observations and calibration data. \citet{redd66} originally placed the association at a distance of 2.1~kpc, while \citet{mass91} estimated a nearer distance of $\sim$1.75~kpc (DM=11.20) following their spectroscopic study of massive stars in the association. The recent recalibration of OB star absolute magnitudes led \citet{hans03} to determine an average distance of 1.2~kpc (DM=10.44) but concluded based on inconsistencies with current stellar wind theory that this was too low and suggested a value of 1.45~kpc (DM=10.80). This is the value we use in this work, which is in agreement with the parallax distance of $1.40 \pm 0.08$~kpc for the Cygnus~X complex \citep{rygl12}.

The uncertainties in our values of $T_{eff}$ and $L_{bol}$ were derived from MC simulations. For the former we use the quoted uncertainties in $T_{eff}$ from \citet{mart05}, \citet{trun07}, and \citet{hump84}, while for the latter we take into account the uncertainties in the observed photometry, derived extinctions, BCs \citep[$\sigma = 0.1$~mag for O-type stars in the $K$-band,][and assumed to be $\sigma = 0.2$~mag for B-type stars]{mart06}, and spectral classification. For the distance to Cyg~OB2 we assume that the distance of 1.45~kpc is correct (small variations in this value up to a few hundred parsecs do not significantly affect the results of this paper, though larger values do).

\subsection{The Hertzsprung-Russell diagram}

\begin{figure*}
\begin{center}
\includegraphics[height=495pt, angle=270]{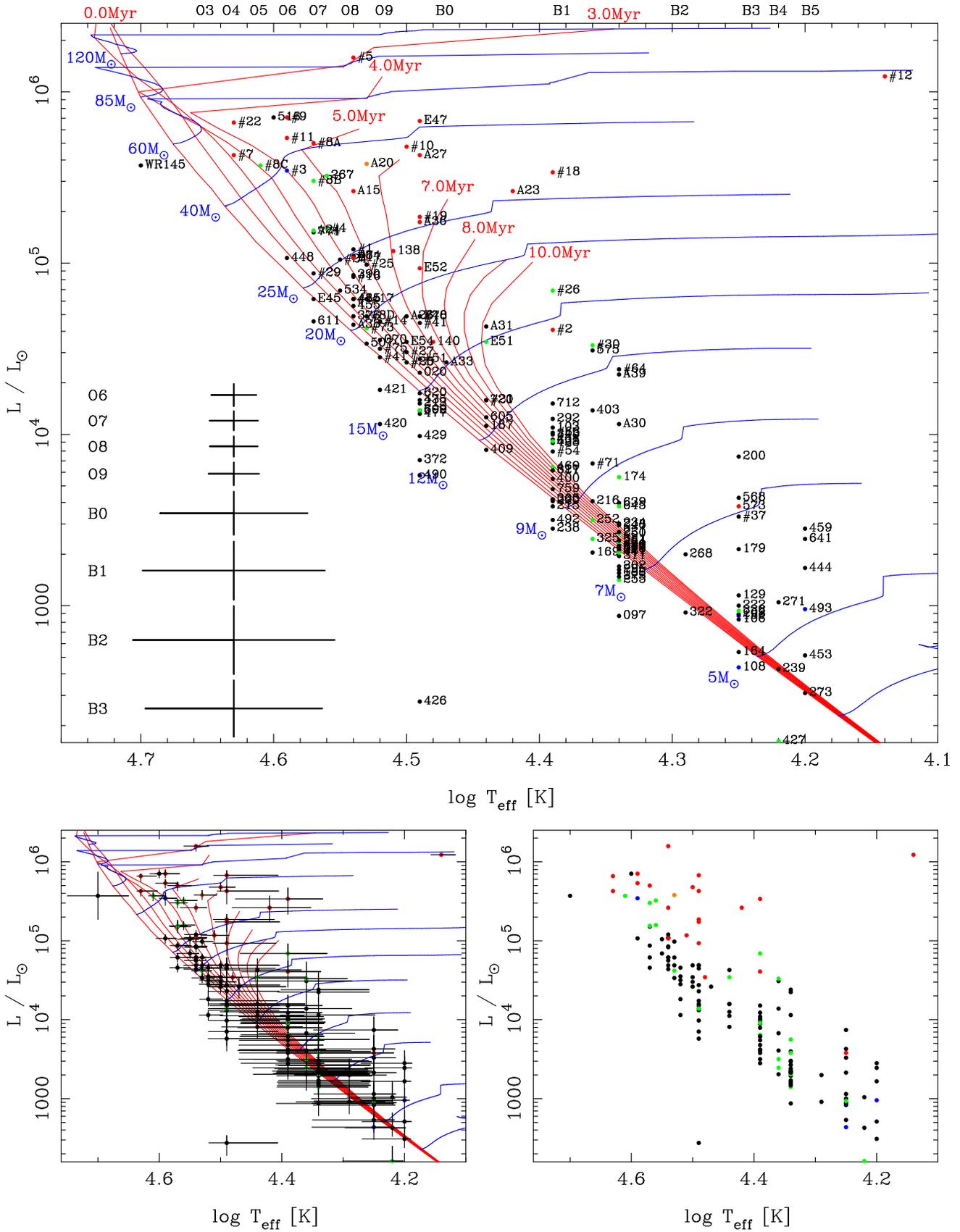}
\caption{HR diagram for Cyg~OB2 showing all the stars compiled in this work as per Figure~\ref{HRD_lejeune}, with the difference that the main panel shows isochrones and evolutionary tracks for rotating ($v_{ini} / v_{crit} = 0.4$) models from \citet[][smoothed for clarity]{ekst12} with the ages and masses labelled. The bottom left figure shows the error bars for each source and the isochrones and evolutionary tracks. The bottom right figure shows the distribution of all stars without labels and evolutionary tracks.}
\label{HRD}
\end{center}
\end{figure*}

The new HR diagram for Cyg~OB2 showing 167 stars is shown in Figure~\ref{HRD_lejeune} compared to the isochrones and evolutionary tracks from the non-rotating models of \citet{leje01} and in Figure~\ref{HRD} compared to the rotating models of \citet{ekst12}. The non-rotating models at a metallicity of $Z = 0.02$ are shown partly for comparison with previous studies \citep[e.g.,][]{hans03,come12}, and also to contrast with the more recent, rotating models at the revised solar metallicity of $Z = 0.014$. An indication of the uncertainties on $L_{bol}$ and $T_{eff}$ are shown in Figures \ref{HRD_lejeune} and \ref{HRD} for each spectral subclass (where possible) and in Figure~\ref{HRD} for individual stars to illustrate the spread they induce in the HR diagram. For O-type stars the uncertainties are typically small, $\sim$0.02~dex in log~$T_{eff}$ and $\sim$0.1~dex in log~$L / L_\odot$, though for the B-type stars the errors rise significantly to $\sim$0.07~dex in log~$T_{eff}$ and $\sim$0.2~dex in log~$L / L_\odot$ due to classification uncertainties and the larger difference in log~$T_{eff}$ between spectral subtypes.

As noted by previous authors \citep[e.g.,][]{mass91,hans03} there is a well-populated main sequence extending up to O5.5V\footnote{The O5V star A37 falls outside of our survey area and therefore is not included in this study, and may not actually be a member of Cyg~OB2, as suggested by \citet{hans03}.}, for which there is now the data to extend it down to mid-B spectral types.

\subsection{A large luminosity spread amongst the B stars}

Many evolved massive stars are evident, both at masses $M > 25~M_\odot$ where such evolution can occur within 2-10~Myr, and at lower masses where evolved B-type stars occupy positions in the HR diagram suggesting ages $> 10$~Myrs. \citet{hans03} noted in her sample a large luminosity spread in the B1-2 dwarfs and this is also apparent here, although it is now more concentrated towards the main-sequence (most likely because the massive stars discovered recently have been fainter as the depth of spectroscopic observations has increased). \citet{hans03} and \citet{negu08} have suggested that their position in the HR diagram may be due to them being foreground contaminants from a separate population. However we note that for their luminosities to be consistent with foreground, main-sequence B-type stars would require some of the most over-luminous stars to lie at distances as low as $\sim$500~pc, which would be inconsistent with their high extinctions (the five most over-luminous B-type stars all have $A_V > 5.5$~mag) and the estimated distance to the extinguishing Cygnus rift of 800-1400~pc \citep{guar12,sale14}.

Alternatively the apparently large luminosity spread amongst the B-type stars could be due to the large uncertainties in $L_{bol}$ and $T_{eff}$ at these spectral types that are smearing out the main sequence in this range. Observational uncertainties in spectral type combined with the steep relationship between $T_{eff}$ and spectral subtype in this range creates large uncertainties in $T_{eff}$ amongst the early B-type stars. Comparing the observed spread in the HR diagram with our estimated uncertainties suggests there is no significant evidence favouring a foreground population of B-type stars or a co-spatial population of mature ($>$10~Myr) stars to explain the observed spread. This argument is supported by the fact that the seemingly high-luminosity B-type stars have a similar spatial distribution to the low-luminosity B-type stars, suggesting they are unlikely to be a separate population.

\subsection{Possible foreground and background contaminants}

A small number of stars fall significantly below the main sequence in the HR diagram such as MT170 (B5V, log~$L/L_\odot = 1.74$), MT426 (B0V, log~$L/L_\odot = 2.51$), and MT427 (B4III, log~$L/L_\odot = 2.19$). MT170 and MT427 appear to be background sources, most likely at distances of $\sim$ 2.19--3.5~kpc if they are on the main-sequence \citep[a number of H{\sc ii} regions and massive young stellar objects have been identified at similar distances by][]{xu13}. Their moderate extinctions ($A_V = 3.6$ and 3.8~mag, respectively) are acceptable given that they are projected against the `reddening hole' in the north-west of the association \citep[see also][who find evidence of a population for A-type stars at such distances in this direction]{drew08}.

The low luminosity of MT426 is harder to explain as it would imply that the star is $\sim$10~kpc distant yet has an extinction of only $A_V \sim 4$ whilst being projected against the centre of the association where the extinction is typically $A_V = 6$--7~mag (Figure~\ref{extinction_voronoi}). MT426 is listed by \citet{mass91} as having `blended' photometry (see Appendix~\ref{s-blended}) with $V = 14.05$~mag and $B-V = 1.95$~mag. These values have led previous authors to derive reasonable extinctions and luminosities for this star, yet such values would imply a $K_s$-band magnitude of $\sim$8.4, yet the source is not detected by 2MASS. Inspection of the UVEX and UKIDSS images reveals a second point source that is highly blended in the lower spatial resolution \citet{mass91} and 2MASS images with its neighbour MT425. The UVEX and UKIDSS photometry, which we have used, does not appear discrepant and we believe it to be sound. It is possible that the spectral type of B0V is erroneous, given the blending reported by \citet{mass91} and the fact that those authors report an identical spectral type for its neighbour \citep[though][report the same spectral type, although it may also have suffered blending]{kimi07}.

Due to the unresolved nature of MT426 and the fact that MT170 and MT427 appear to be background sources we argue that all three are non-members of Cyg~OB2 and exclude them from further analysis of the association. They are noted in Table~\ref{obstars} as `Unresolved' or `Background' sources, and physical properties are not derived for them.

Finally we note that the membership status of Cyg~OB2~\#12 has been questioned in the past because of its advanced evolutionary state compared to the early O-type stars in the association and its high luminosity \citep{walb73}, with the suggestion that the star might be a foreground, though it could also be an luminous blue variable \citep{clar12} or a similar-luminosity binary \citep[][recently identified a companion to the star, but note that it is too faint to affect the calculated properties of the star]{caba14}.

\subsection{Physical quantities derived from the HR Diagram}
\label{s-physical}

Stellar ages and initial masses were calculated based on the position of each star in the HR diagram relative to the non-rotating $Z = 0.02$ models of \citet[][commonly used for previous studies of Cyg~OB2]{leje01} and the rotating $Z = 0.014$ \citep[the revised solar metallicity value,][]{aspl05} models of \citet{ekst12}, both from the Geneva group. The rotating models have an initial rotation rate of $v_{ini} / v_{crit} = 0.4$, which corresponds to typical main sequence velocities of 110-220~km\,s$^{-1}$ for massive stars \citep{ekst12}, in agreement with observations \citep[e.g.,][]{duft06}. However it is not currently clear whether stars are born with a narrow or broad range of initial rotation velocities.

Stellar mass loss is an important ingredient in stellar models, highly influenced by the metallicity \citep{meyn94}, rotation rate \citep{maed00}, and the exact mass-loss prescription used \citep[e.g.,][]{vink01}. The non-rotating evolutionary models of \citet{leje01} used the radiative mass-loss rates from \citet{deja88} for stars in the blue part of the HR diagram, while \citet{ekst12} update this with the prescription from \citet{vink01} where possible, and also take into account supra-Eddington and mechanical mass-loss processes in their models. Different mass-loss rates can play a significant role in the evolution of massive stars, their positions in the HR diagram, and therefore the physical parameters derived in such a way. These differences are most pronounced for red supergiants and the most massive stars \citep[40--120~M$_\odot$,][]{ekst12}, with the latter having considerably increased main-sequence lifetimes when mass-loss rates are increased \citep{meyn94}. For moderately massive stars (20--40~M$_\odot$) the effects of mass-loss can be less notable \citep[e.g.,][note that the main-sequence lifetime of 25~M$_\odot$ stars increased by only 5\% when the mass-loss rate was doubled]{meyn94}.

The ages and masses derived from these models are listed in Table~\ref{obstars} with uncertainties. For the vast majority of objects this is a simple process and we do this on the assumption that all the stars are on their first journey across the HR diagram from hotter to cooler temperatures (generally a valid assumption since stars move back to hotter temperatures much faster and therefore it is rarer to observe stars during this transition). For stars that fall to the left of the zero-age main-sequence in the HR diagram we assume that their positions are due to uncertainties in their luminosity or effective temperature and that these are zero-age main-sequence objects.

To derive uncertainties on the age and initial mass of stars we run a MC simulation that takes into account both the observational uncertainties on $L_{bol}$ and $T_{eff}$ as well as uncertainties due to unresolved binary companions and an intrinsic spread in both luminosity and temperature for a star of a given initial mass and age. The latter simulates the spread in the HR diagram at a given initial mass and age caused by variations in fundamental parameters such as rotation and metallicity, and was estimated to be 0.12~dex in log~$L / L_\odot$ and 0.02~dex in log~$T_{eff}$ from inspection of the spread in the upper main-sequence in rotating stellar models, and an extrapolation of the spread measured by \citet{houk97} in Hipparcos data. The effects of unresolved secondaries were modelled in our MC simulation using the binary characteristics found by \citet{kimi12b} for Cyg~OB2: a binary fraction of 90\% for O-type stars (we assume a slightly lower binary fraction of 80\% for the B-type stars) and a flat mass ratio distribution over the range $q = 0.005$--1.0. The added luminosity uncertainty due to unresolved binary companions is generally small, typically adding an uncertainty of 0.1~dex, and therefore we do not consider more complex multiple systems. The uncertainties shown in Figures \ref{HRD_lejeune} and \ref{HRD} (for both individual stars and representative values for each spectral subtype) are the uncertainties on the calculated values of $L_{bol}$ and $T_{eff}$ only and do not take into account these additional uncertainties used when calculating the stellar mass and age. The median fractional uncertainty on the stellar mass is found to be $\sigma_M / M \sim 0.14$ and for stellar age is $\sigma_\tau / \tau \sim 0.3$ (for stars in the range 0--10~Myr).

\section{The age of Cyg~OB2}
\label{s-ages}

\begin{figure*}
\begin{center}
\includegraphics[height=495pt, angle=270]{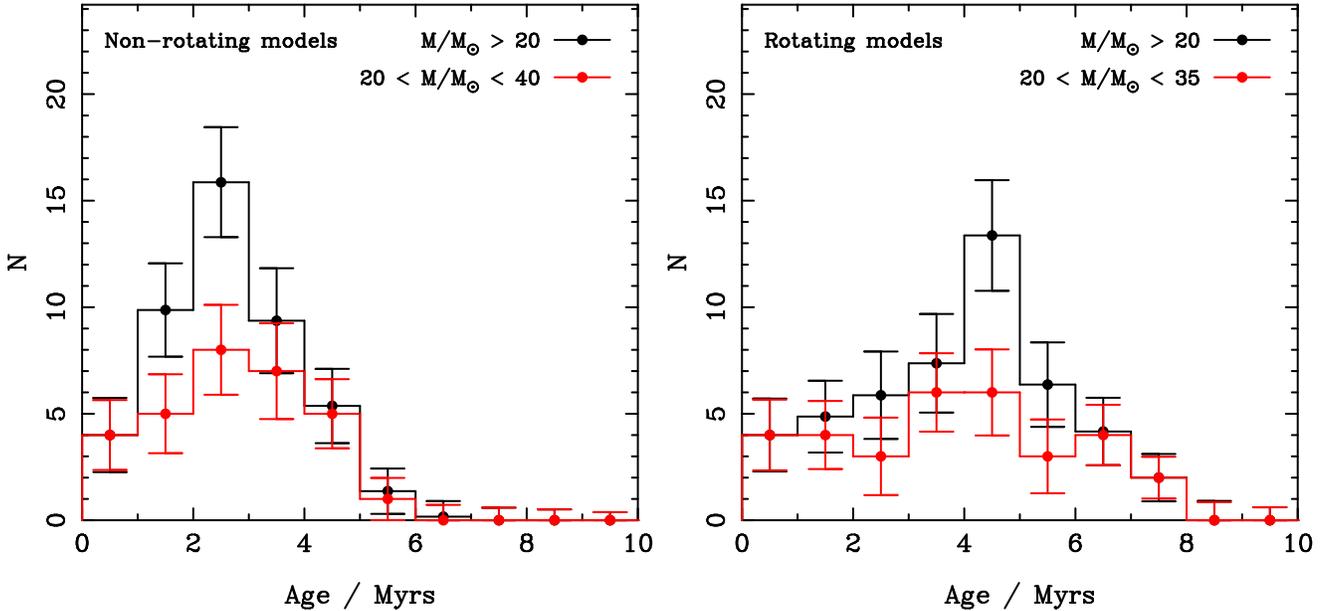}
\caption{Histograms showing the distribution of derived ages for stars with reliably determined ages from the non-rotating (left) and rotating (right) models. The ages of all stars with $M \geq 20$~M$_\odot$ are shown with a black line. In the left panel we also show in red the age distribution of stars with $20 \leq M / M_\odot \leq 40$, which should be complete and unbiased in the age range of 0--4.5~Myr. In the right panel we show in red the age distribution of stars with $20 \leq M / M_\odot \leq 35$ that are unbiased in the age range of 0--6~Myr. 1$\sigma$ uncertainties are shown for all histograms derived from a Monte Carlo simulation of the data analysis and binning process.}
\label{age_histogram}
\end{center}
\end{figure*}

Using the stellar ages derived from the HR diagram we can assess the age distribution of stars in Cyg~OB2 and hence the star formation history of the region. This is shown in Figure~\ref{age_histogram} using both the non-rotating and rotating evolutionary models to highlight the different age distribution that they each imply. We only show the 47 stars more massive than 20~M$_\odot$ (down to a main-sequence spectral type of $\sim$B0.5V) based on our estimated completeness level and because the positions of stars less massive than this in the HR diagram do not provide an accurate indication of their ages (due to the closeness of the isochrones and the large uncertainty on $T_{eff}$ for these stars). This excludes all the apparently over-luminous B-type stars discussed earlier. The uncertainties shown in Figure~\ref{age_histogram} were calculated from a Monte Carlo simulation of the data analysis and binning process.

\subsection{Comparison of age distributions from rotating and non-rotating models}

The age distribution of massive stars derived from the non-rotating models shows a strong peak at 2-3~Myrs with the majority of stars in the age range of 0-5~Myrs, in good agreement with previous studies of the massive star population of Cyg~OB2 \citep[e.g.,][]{hans03,come12}. However, since a 20~M$_\odot$ star has a lifetime of only $\sim$8~Myr \citep[in the][models]{leje01} then our sample of massive stars with masses $>$20~M$_\odot$ is only sensitive to stars $<$8~Myrs old. We therefore cannot reliably infer the star formation history greater than 8~Myrs ago when using these models. Furthermore this distribution is biased by the fact that the most massive stars have total lifetimes $<$4~Myrs. For example a 60~M$_\odot$ star is estimated to have a total lifespan of $\sim$3.5~Myrs, and therefore any such stars born more than 4~Myrs ago will not be seen now. To rectify this well-known bias we also show in Figure~\ref{age_histogram} the distribution of ages of stars in the mass range $20 \leq M / M_\odot \leq 40$, which should be unbiased in the age range of 0-4.5~Myr. This presents a slightly different picture with a smaller peak suggesting that this peak age is an observational bias caused by the short lifetimes of the most massive members of the association.

The rotating stellar models of \citet{ekst12} (Figure~\ref{age_histogram}) show a different picture with a prominent but wide peak at an age of 4-5~Myr. The increase in the typical age of massive stars in Cyg~OB2 when using the rotating stellar models can be attributed to the changes in evolutionary timescales induced by rotation, which increases the main sequence lifetime by $\sim$25\% \citep{ekst12}. The marked influence of this on stellar ages is evident in the two HR diagrams (Figures \ref{HRD_lejeune} and \ref{HRD}). For example, the group of O4-7 supergiants with $L / L_\odot \sim 5 \times 10^5$~L$_\odot$ have ages of 1-3~Myr according to the non-rotating models, but have ages of 3-5~Myr from the rotating models. The older ages are in better agreement with those derived from lower-mass stars in Cyg~OB2, e.g., \citet{drew08} estimate an age of 5--7~Myrs from A-type stars, while \citet{wrig10a} derive an age of 3.5--5.25~Myrs from an X-ray selected sample of approximately solar-mass stars.

The increased stellar lifetimes of rotating stellar models means that the mass range over which our sample is unbiased has also changed relative to the non-rotating models. Figure~\ref{age_histogram} shows the age distribution of stars in the mass range $20 \leq M / M_\odot \leq 35$, which should be complete in the age range 0-6~Myrs. This distribution is remarkably flat over the unbiased age range, suggesting a relatively constant level of star formation over the last 6~Myrs. The peak stellar age previously observed at 4-5~Myrs is mostly due to stars in the HR diagram (Figure~\ref{HRD}) at masses of 40--60~$M_\odot$ (the O4-7 supergiants noted earlier) that have maximum stellar ages of $<$5~Myrs. The large number of slightly lower mass (25--50~$M_\odot$) stars with ages of 5--6~Myrs (the O7-9 supergiants observed in the HR diagram) could suggest that many of the more massive members have already evolved to their end states and their loss biases the observed age distribution.

The comparisons with both sets of evolutionary models indicate a number of stars in Cyg~OB2 with very young ages, $\leq$1~Myr. The existence of such young stars is not impossible but appears unlikely given that long wavelength images \citep[e.g., Figure~\ref{map}, and][]{schn06} show that the majority of dust and molecular gas been evacuated from the cavity in which Cyg~OB2 resides. Furthermore both \citet{wrig10a} and \citet{guar13} found a paucity of Class~I or heavily embedded low-mass stars within the association, suggesting the low-mass stellar population is not so young \citep[though this could also be due to the high levels of feedback in the vicinity of the association,][]{wrig12a,guar14}. The typical uncertainties in the HR diagram shown in Figures \ref{HRD_lejeune} and \ref{HRD} suggests that the positions of these stars are not incompatible with them being slightly older ($>$1--2~Myr) and the lack of very massive stars ($>$30~M$_\odot$) with ages $<$1~Myr (where the large separation of isochrones allows young ages to be estimated more accurately) supports this idea.

To conclude, our unbiased age distribution derived from rotating stellar models suggests a relatively constant level of star formation over the past $\sim$6~Myr (Fig~\ref{age_histogram}) with some evidence for a peak in the star formation intensity between 4--5~Myr ago. We cannot be certain from this sample when star formation in the region began, but there is some evidence that it started at least $\sim$6--8~Myr ago. Addressing the question of when star formation ceased is harder, but our best estimate is that it ended $\sim$1~Myr ago. As will be discussed in Section~\ref{s-cmf} these results are in approximate agreement with our star formation history fits to the mass function of stars in Cyg~OB2. The use of different mass-loss prescriptions is not expected to significantly affect the evolution of stars in our `unbiased' mass range of 20--35~M$_\odot$ \citep{meyn94}, and therefore this is unlikely to explain the age spread observed. It is possible that the large spread in the HR diagram could instead be due to a single-aged population with an extremely large range of rotation velocities, which has been shown to lead to large spreads in the HR diagram \citep{brot11} and could reproduce the entire spread seen in our HR diagram. However, for this to be the case the stars nearest the zero aged main-sequence in the HR diagram would have to be rotating very rapidly with speeds of 400-450~km/s, including Cyg~OB2 \#20 and \#29, for which \citet{herr99} measured $V \, \mathrm{sin} \, i = 25$ and 180~km/s, respectively. This would therefore require Cyg~OB2 \#20 to have been observed virtually pole-on, which makes this scenario unlikely, though further high resolution spectroscopy of other stars in Cyg~OB2 would be useful to test this hypothesis in more detail. It is also worth noting though that this scenario is not supported by measurements of rotation rates in other clusters \citep[e.g.,][]{duft06}.

The uncertainties in the stellar ages we have calculated could explain some of this spread, but do not appear to be sufficiently large enough to explain the entire $\sim$5~Myr spread. Furthermore the age distribution of our `unbiased' sample shown in Figure~\ref{age_histogram} does not resemble a single-aged population broadened by observational uncertainties (which should have a more centrally-concentrated peak) and does appear to represent a spread of ages amongst the OB star population.

\subsection{Implications for large age spreads in OB associations}

Large age spreads have been reported in many star clusters and associations, typically diagnosed from luminosity spreads in colour-magnitude diagrams for low-mass pre-main sequence stars. Some authors have suggested that such spreads in luminosity do not equate to age spreads and may instead be due to intrinsic luminosity spreads amongst a coeval population \citep[e.g.,][]{hill08,jeff11b} potentially caused by different accretion histories \citep[e.g.,][]{bara09}. Age spreads diagnosed from post-main sequence massive stars are not believed to suffer from such issues since the intrinsic spread and measurement uncertainties in $L_{bol}$ and $T_{eff}$ are typically smaller than the separation of isochrones \citep[see also][]{mass03}. Therefore if the HR diagram is sufficiently well sampled to overcome statistical and observational uncertainties (as it is here), age spreads derived from high-mass post-MS stellar evolution models can provide more accurate age diagnostics than low- and intermediate-mass pre-MS stellar evolution models.

The reported age spreads for low-mass stars are often argued as representing a continuous period of star formation within a single cluster or association and in a small area of space \citep[e.g.,][]{pall00}. Cyg~OB2 however occupies a much larger region than typical young compact star clusters such as the Orion Nebula Cluster, with the stars spread over $>$10~pc. It has recently been argued that Cyg~OB2 has always been a large and low density OB association and was never a dense and compact star cluster \citep[][see also \citealt{park14})]{wrig14b}. It is therefore possible that the observed age spread is due to either an age gradient or a series of discrete star forming events that have since mixed. This would imply that Cyg~OB2 was not born as a single star cluster but as a `distribution' of smaller groups or clusters of stars with a range of stellar ages. We searched for evidence of a spatial variation in the stellar age distribution to support this theory, but could not find any such evidence. If the association is composed of multiple populations of different ages then they are well mixed.

\section{The upper end of the IMF in Cyg~OB2}

\begin{figure*}
\begin{center}
\includegraphics[height=495pt, angle=270]{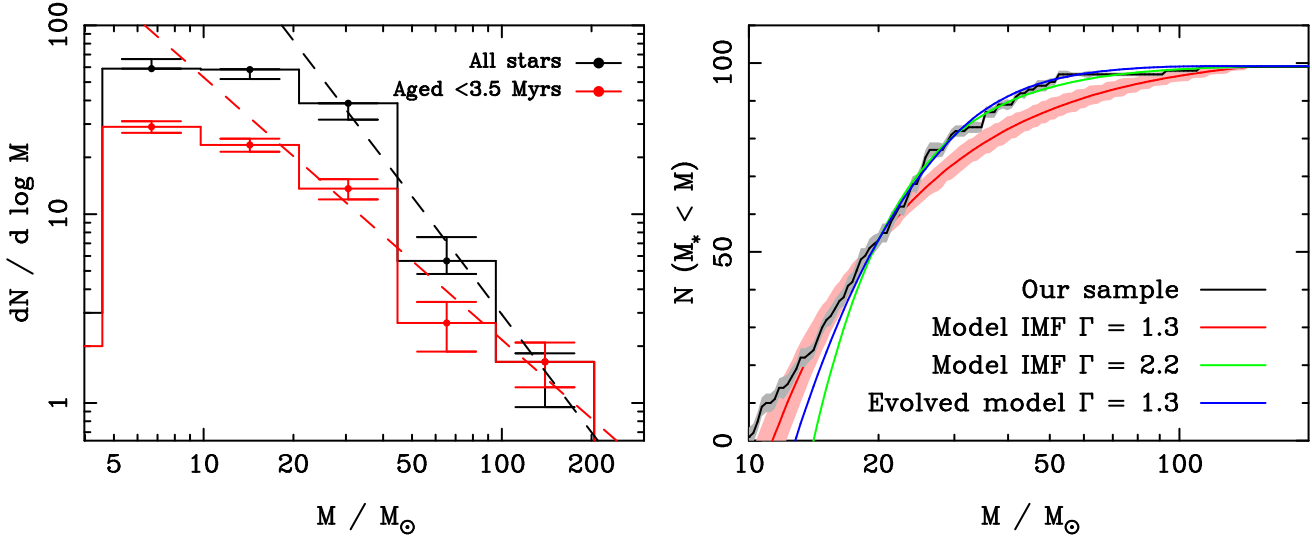}
\caption{Mass functions for massive stars in Cyg~OB2 derived using the rotating stellar models of \citet{ekst12}. 
{\it Left:} Histograms showing the distribution of derived masses for all 167 stars in our sample (black histogram, with 1$\sigma$ error bars determined from MC simulations) and for an age-unbiased sample of 82 stars aged $<$3.5~Myrs (red histogram, with 1$\sigma$ error bars). Both histograms have been fitted using an ordinary least squares regression of $Y$ on $X$ \citep[OLS ($Y|X$),][]{isob90} for the mass bins $>$20~M$_\odot$, giving slopes of $\Gamma = 2.06 \pm 0.06$ and $\Gamma = 1.39 \pm 0.19$ for the complete and age-unbiased samples respectively (note that these uncertainties do not take into account random sampling of the IMF). 
{\it Right:} A cumulative mass function showing $N \, (M_\star < M)$ as a function of $M$ for all 98 stars in our sample more massive than 10~$M_\odot$ (black line) with the 1$\sigma$ uncertainty (calculated from MC simulations) shown in grey. Also shown are the results of simulated IMFs for $\Gamma = 1.3$ (red line, with 1$\sigma$ random sampling uncertainty shown in pink) and $\Gamma = 2.2$ (green line) calculated using the equations of \citet[][see full explanation in the text]{masc13}. The blue line shows a simulated and evolved IMF for $\Gamma = 1.3$ for stars aged 3--7~Myr (i.e. a constant star formation rate from 3 to 7 Myr ago, and no star formation since then). The uncertainties for three simulated IMFs are similar and so only one is shown for clarity.}
\label{mass_histogram}
\end{center}
\end{figure*}

In this section we analyse the IMF of the massive stars in Cyg~OB2, as derived using rotating stellar models \citep[because][find that such masses are in much better agreement with dynamical masses than those derived from non-rotating stellar models]{weid10}. Figure~\ref{mass_histogram} shows a histogram of the masses of all 167 stars in our sample, showing a rapid increase as one goes from high to low masses, with the data flattening off for stars less massive than our estimated completeness limit of $\sim$20~M$_\odot$. To calculate the power-law slope of the mass function we fitted this data using an ordinary least squares regression of $Y$ on $X$ \citep[OLS ($Y|X$),][]{isob90} of the form $dN / d \, \mathrm{log} \, m \propto m^{-\Gamma}$, and for the mass bins $>$20~M$_\odot$. Because the uncertainty on this fit is just a fitting uncertainty we also performed a MC simulation to determine the true uncertainty, varying the masses of the individual stars according to their uncertainties and adjusting the number of stars in each bin as appropriate. The 1$\sigma$ spread in the best fitting slope that results is then combined with the fitting error to produce the true uncertainty on the mass function slope. The best fitting mass function slope for all stars in our sample is $\Gamma = 2.06 \pm 0.06$, significantly steeper than the canonical Salpeter slope of $\Gamma \sim 1.35$ \citep{salp55}, although it agrees with that determined by other authors for the massive stars in Cyg~OB2 \citep[e.g.,][]{kimi07,come12}. Note however that this uncertainty is still falsely low because it does not take into account the random sampling of the IMF.

The mass distribution of stars in our sample is likely to be altered by the loss of the most massive stars that have already evolved to their end states (Section~\ref{s-ages}), steepening the observed mass function. To counter this we have attempted to identify an unbiased sample of stars from which to study the initial mass function in Cyg~OB2, selecting all stars with ages $<$3.5~Myrs, which we estimate from the rotating stellar models (Figure~\ref{HRD}) as the age at which the most massive stars in Cyg~OB2 will have evolved to their end states. The mass function of this unbiased sample of 82 stars is shown in Figure~\ref{mass_histogram}, which shows a slightly shallower slope to that of the full sample. An OLS ($Y|X$) fit gives a slope of $\Gamma \sim 1.39 \pm 0.19$, in good agreement with the concept of a universal IMF at high masses with a slope of $\Gamma = 1.3$ \citep{krou01,bast10}. This value also agrees with the mass function measured at lower masses \citep[$\Gamma = 1.09 \pm 0.13$,][]{wrig10a}, and the mass function of massive stars found by \citet{mass91} of $\Gamma = 1.0 \pm 0.3$, although their sample was considerably smaller than the current sample and may therefore have been incomplete.

\subsection{The Cumulative Mass Function}
\label{s-cmf}

To avoid the well-known and negative effects of binning our data, and to overcome any uncertainties arising from the narrow age range used for the unbiased mass sample we have also considered the mass function of stars in Cyg~OB2 represented as a cumulative function (CF). This is shown in Figure~\ref{mass_histogram} with 1$\sigma$ uncertainties calculated from a MC simulation using the full uncertainties on all stellar masses (the 1$\sigma$ uncertainties shown are very small due to the cumulative nature of the function and the lack of binning).

The CF shows a steady increase up to $\sim$30~M$_\odot$, and then flattens off, reaching a total of 98 stars with masses $> 10$~M$_\odot$. We compare our CF with simulated IMFs generated using the equations described in \citet{masc13}, varying $\alpha$ ($= \Gamma + 1$), but fixing $\beta = 1.4$ (the low-mass exponent has no effect on the high-mass IMF, but is required for the calculations) and drawing individual stellar masses from the mass range 0.01--150~M$_\odot$. We continue to do this until the number of stars with $M > 20$~M$_\odot$ reaches the observed number of 98 and use this to produce a cumulative function scaled to the observed number of stars with $M < 20$~M$_\odot$. The simulated IMF is generated 10,000 times and the median CF shown in Figure~\ref{mass_histogram} with $\pm$1$\sigma$ range illustrated.

We initially simulated an unevolved (i.e. counting all stars) `universal' IMF \citep{krou01} with $\Gamma = 1.3$, but found that this over-predicted the number of very massive stars ($> 50$~M$_\odot$) relative to the moderately massive stars (20--50~M$_\odot$) and we can reject a standard, unevolved mass function of this type with a confidence of $\sim$3$\sigma$. Varying $\Gamma$ we found a best fit with an un-evolved IMF for $\Gamma = 2.2^{+0.6}_{-0.3}$, in approximate agreement with that found fitting a binned IMF for our entire sample (but with larger uncertainties that also take into account the random sampling of the IMF).

We then simulated a number of evolved mass functions using the `universal' $\Gamma = 1.3$ IMF and different star formation histories. For a star of a given mass and age we compute its position in the HR diagram according to the models of \citet{ekst12} and remove those stars that have evolved beyond their reddest position in the HR diagram. This excludes WR stars from the simulated mass functions, matching our observed CF that also leaves them out (because of the difficulty of determining their initial masses).

We found that a number of different star formation histories provided satisfactory fits (within 1$\sigma$) to the observed mass function, preventing us from accurately determining it in this way, though we were able to place some partial constraints on it. We find fits using a `universal' $\Gamma = 1.3$ IMF requires the star formation rate over the last $\sim$3~Myr to be very low, $<$20\% of the total star formation in Cyg~OB2, with the fit improving slightly as the fraction of star formation in this time period decreases. Exploring models with this feature we find acceptable fits when the duration of star formation is at least 2~Myrs (i.e., a constant level of star formation from 3 to 5~Myrs ago), with the best fit obtained for a model with a constant star formation rate over the time period of 3 to 7~Myr ago, in reasonable agreement with that found from the distribution of stellar ages studied earlier. Significant star formation prior to $\sim$7~Myr ago degrades the fit to the cumulative mass function. These results are in good agreement with the stellar age distribution discussed in in Section~\ref{s-ages}.

To conclude, the current mass function of massive stars in Cyg~OB2 is relatively steep, with a slope of $\Gamma \sim 2.1$, but the stellar age distribution suggests that it has been steepened by the most massive stars evolving to their end states. From both an unbiased sample of stars and a modelling an evolved cumulative mass function we find that the mass function is consistent with the canonical $\Gamma = 1.3$ `universal' IMF \citep{salp55,krou01}.

\subsection{Have their been any supernovae in Cyg~OB2?}

There is considerable evidence, both in the age distribution of stars and in the steepened IMF, to suggest that some of the most massive stars in Cyg~OB2 have already evolved to their end states and therefore that there has already been a supernova within the region.

Despite this there are not any known supernova remnants (SNRs) within Cyg~OB2, although \citet{gree09} do list a number of SNRs in the wider Cygnus~X region (including $\gamma$-Cygni, G78.2+2.1) but these are predominantly to the west in the Cyg~OB1 and OB9 associations. \citet{butt03} identified an expanding shell of molecular and dusty material from CO$J = 1 \rightarrow 0$ and {\it IRAS} observations that they suggested could be due to a SNR, but this cavity could equally arise from the combined effects of ionising radiation and stellar winds from the massive stars in Cyg~OB2. The lack of observed SNRs in Cyg~OB2 does not however rule out any previous supernovae in the association, since SNRs typically remain visible for only $\sim$10,000~yrs \citep[e.g.,][]{leve98} and may not leave any detectable signature when expanding into hot, low-density media such as the cleared-out cavities in OB associations \citep{chu97}.

There have been a number of very high-energy sources identified towards Cyg~OB2 that have been considered as possible compact stellar remnants \citep[e.g.,][]{bedn03,butt06}, but confirming their nature or identifying counterparts at other wavelengths has proved difficult. \citet{abdo09} discovered a $\sim$115,800~yr old $\gamma$-ray pulsar (J2032+4127) with a frequency of 6.98~Hz toward Cyg~OB2, which has since been detected in the radio with the same frequency \citep{cami09} and is also believed to be associated with the previously-unidentified TeV source J2032+4130 \citep{ahar02} as its pulsar wind nebula. \citet{cami09} argue that the pulsar is likely to be located within the Cyg~OB2 association, and combined with our newly derived age of $\sim$1--7~Myrs this is consistent with the pulsar originating within Cyg~OB2 supporting the theory that the association has already seen it's first supernova.

Another indication of past supernovae is the presence of high velocity `runaway' OB stars that are ejected when their binary companion explodes as a supernova \citep{blaa61}. \citet{come07} identified an O4If runaway star with a proper motion of $\sim$40~km/s moving away from Cyg~OB2 and argued that the star was likely ejected from the association $\sim$1.7~Myrs ago. The authors estimate a mass of $70 \pm 15$~M$_\odot$ and an age of $\sim$1.6~Myrs (by comparisons with non-rotating stellar isochrones) suggesting the star was ejected from Cyg~OB2 shortly after it was born, and therefore not from a supernova explosion. The other mechanism for stellar ejection is via dynamical encounters in dense stellar systems \citep{pove67}, yet this process requires much higher stellar densities than are believed to have existed in Cyg~OB2 \citep{wrig14b} making it unlikely. Using the rotating stellar models of \citet{ekst12}, and with updated photometry we re-calculated the properties of this star. We find $B-K_s = 5.27$~mag \citep[using Tycho-2 and 2MASS photometry,][]{hog00,cutr03}, which for a star of type O4If with $(B-K_s)_0 = -1.15$~mag \citep{mart06} gives $A_V = 5.22$~mag (using the reddening law determined above) and therefore log~$L_{bol} = 5.85$~L$_\odot$ and $T_{eff} = 40,422$~K. This implies an initial mass of $\sim$55~M$_\odot$ and an age of 4-5~Myr. This would suggest the star was ejected from Cyg~OB2 when it was about $\sim$3~Myr old, the age at which the most massive star in a population might explode as a supernova. This is therefore consistent with a scenario whereby the star's very massive ($>$100~M$_\odot$) companion exploded as a supernova and ejected it's companion from the association. The remnant of the massive star was presumably ejected from Cyg~OB2 in the opposite direction and could be searched for to verify this scenario.

\section{The total mass of Cyg~OB2}

Using a sample of massive stars that is believed to be complete within a given mass range, and the assumption that the IMF has a `universal' form in all star clusters and OB associations, it is possible to estimate the total mass of a group of stars beyond that which have been detected. To do this for Cyg~OB2 we have selected a mass range of 20--40~$M_\odot$ for which we believe our sample is complete (at least for stars aged $<$6~Myrs). Above 40~M$_\odot$ our sample will be incomplete for stars aged $<$6~Myrs and below 20~M$_\odot$ we have argued that our sample is likely to be observationally incomplete. Based on the stellar masses determined above the number of stars in this mass range is $36^{+1}_{-4}$ (uncertainties calculated from MC simulations).

To estimate the total mass of Cyg~OB2 we ran a MC simulation by sampling stellar masses from a `universal' IMF with $\alpha = 2.30$ ($\Gamma = 1.3$) and $\beta = 1.40$ \citep{krou01,masc13} across the entire mass range. We proceeded to count the total mass of stars produced and the number of stars in the mass range of 20--40~M$_\odot$, until the latter number reached the observed number of stars in that mass range. We performed this simulation 10,000 times, recording the total stellar mass that was necessary in each iteration to produce the observed number of stars. Furthermore we varied the observed number of stars, $36^{+1}_{-4}$, in each iteration according to the uncertainties on that number. This MC simulation therefore takes into account both the uncertainty in the total number of stars observed, and the variation in the number of stars in a cluster of a given mass due to the random sampling of the mass function.

We find that the observed number of massive stars in Cyg~OB2 in the mass range of 20--40~M$_\odot$ can be reproduced if the association has a total stellar mass of $16500^{+3800}_{-2800}$~M$_\odot$. This is based on three assumptions, firstly that we are complete in the chosen mass range, as argued above. Secondly, that the vast majority of star formation in Cyg~OB2 occurred in the last 6~Myr (broadly supported by the age distribution and the mass function simulations, though it is difficult to test this assumption without extending the completeness of the sample to lower masses). The final assumption is that the IMF in Cyg~OB2 is well represented by the `universal' IMF of \citet{krou01}. This is supported by our IMF simulations at high stellar masses and that found at lower masses \citep{wrig10a}, however a small variation can lead to large differences in the total stellar mass. For example increasing (decreasing) the high-mass exponent $\alpha$ by 0.1 increases (decreases) the total stellar mass by 5600 (3900) M$_\odot$. However, for comparison with other young clusters and associations that have also adopted such an IMF we believe this is the most appropriate choice.

Our mass estimate is smaller than some previous mass estimates of Cyg~OB2, notably that of \citet{knod00} who estimated a mass of $(4 - 10) \times 10^4$~M$_\odot$ for the association. \citet{knod00} based their estimate on the number of stars brighter than a given magnitude (equivalent to an F3V star) from a background-subtracted 2MASS star counts study. The choice of a background region in this crowded and complex area of the Galactic Plane is fraught with difficulties and could lead to large uncertainties in any resulting star counts or mass estimates. \citet{knod00} also estimate the total number of O stars in Cyg~OB2 to be $120 \pm 20$, an estimate almost double the size of our census and one which other studies have not supported \citep[e.g.,][]{hans03}. Our mass estimate is more consistent with recent estimates based on studies of lower-mass stars, e.g., \citet{wrig10a} estimated $(2 - 4) \times 10^4$~M$_\odot$ from an X-ray study and \citet{drew08} estimated $(1 - 4) \times 10^4$~M$_\odot$ from a photometric study of A-type stars in the association. 

\section{Conclusions}

We have compiled a census of all known massive stars in the nearby OB association Cygnus OB2. We have gathered data from across the literature to achieve a census containing 169 primary OB stars, including 52 O-type stars and 3 WR stars.

We have used the sample to measure the extinction law towards Cyg~OB2, fitting the available photometry to a \citet{fitz07} $R_V = 2.91 \pm 0.06$ extinction law. This is in good agreement with that determined by \citet{hans03} in the optical and differing only in the near-IR due to shifts in the intrinsic near-IR colours of O-type stars in recent years. We use this new law to derive individual extinctions for all the stars in our sample, the majority of which have extinctions $A_V = 4$--7~mag, with a typical uncertainty of $\sim$0.3~mag.

A Hertzprung-Russell diagram was compiled and compared to both non-rotating and rotating stellar evolution tracks, from which stellar masses and ages are calculated. Uncertainties are derived for all quantities using a MC simulation that takes into account all available uncertainties including quantified classification uncertainties, photometric uncertainties, extinction law uncertainties, estimated spreads in fundamental parameters, and unresolved binaries.

The distribution of stellar masses is assessed using both a binned mass function and a cumulative mass function, both of which are affected by the loss of the most massive stars to stellar evolution. The current mass function is relatively steep with a slope of $\Gamma = 2.2$ \citep[in agreement with that found by][]{kimi07}, though when considering either an unbiased sample of stellar masses or by modelling the effects of stellar evolution on the mass function we find it to be consistent with a Salpeter-like `universal' slope of $\Gamma = 1.3$.

The non-rotating stellar evolution models suggest a peak stellar age of 2--3~Myr, as derived by previous authors, while the rotating stellar models suggest a slightly older peak age of 4--5~Myr. The results from the rotating stellar models are favoured, both for physical reasons, and because the age distribution that they imply is in better agreement with that found from studies of low- and intermediate-mass stars. When considering only stars in a mass range unaltered by evolutionary effects the age distribution is not inconsistent with a relatively constant level star formation over at least the past $\sim$6~Myrs. Our star formation history fits to the mass function of stars in Cyg~OB2 suggest star formation most likely began $\sim$7~Myr ago. The lack of molecular material or dust in the immediate vicinity of the associations suggests star formation likely ceased at least $\sim$1~Myr ago, and this is also supported by our mass function models. Our best estimate is that the majority of star formation occurred during the period of 1--7~Myr ago, with a possible in the star formation intensity between 4--5~Myr ago.

Given the evidence from both the age distribution and the mass function that some of the most massive stars have already evolved to their end states we searched the literature for evidence of this and found both a young pulsar and a runaway O star that may have originated in Cyg~OB2, consistent with the view that Cyg~OB2 has already seen its first supernova.

Finally we use a MC simulation to estimate the total mass of the association, taking into account the loss of the most massive stars, and estimate a total stellar mass of $16500^{+3800}_{-2800}$~M$_\odot$ \citep[assuming a][IMF]{krou01} that has formed over the past 6~Myr. This is at the lower end of many previous estimates, based on the number of low- and intermediate-mass stars, but consistent given the uncertainties involved in extrapolating the IMF.

In conclusion our new census of massive stars in Cyg~OB2 provides an updated view of the high-mass stellar population in one of the largest groups of young stars in our Galaxy and for the first time these results show a good agreement with those determined for low- and intermediate-mass stars, thereby providing a coherent picture for this massive OB association. In a future paper we will use new photometry and spectroscopy to search for previously undiscovered massive stars, extending this census to both lower-luminosity B-type stars in Cyg~OB2, and O and B-type stars over a wider area and further along the sightline.

\section{Acknowledgments}

We would like to thank Facundo Albacete Colombo, Nate Bastian, Mario Guarcello, Artemio Herrero, Ian Howarth, Raman Prinja, and Nolan Walborn for useful discussions and comments on this paper. We would like to thank the referee, Matthew Povich, for a helpful and insightful report that improved the content of this paper. NJW was supported by a Royal Astronomical Society research fellowship. This research has made use of NASA's Astrophysics Data System and Simbad and VizieR databases, operated at CDS, Strasbourg, France.

\bibliographystyle{mn2e}
\bibliography{/Users/nwright/Documents/Work/tex_papers/bibliography.bib}
\bsp

\begin{appendix}

\section{UVEX Photometry and transformations to the $UBV$ System}
\label{s-blended}

\begin{figure*}
\begin{center}
\includegraphics[height=500pt, angle=270]{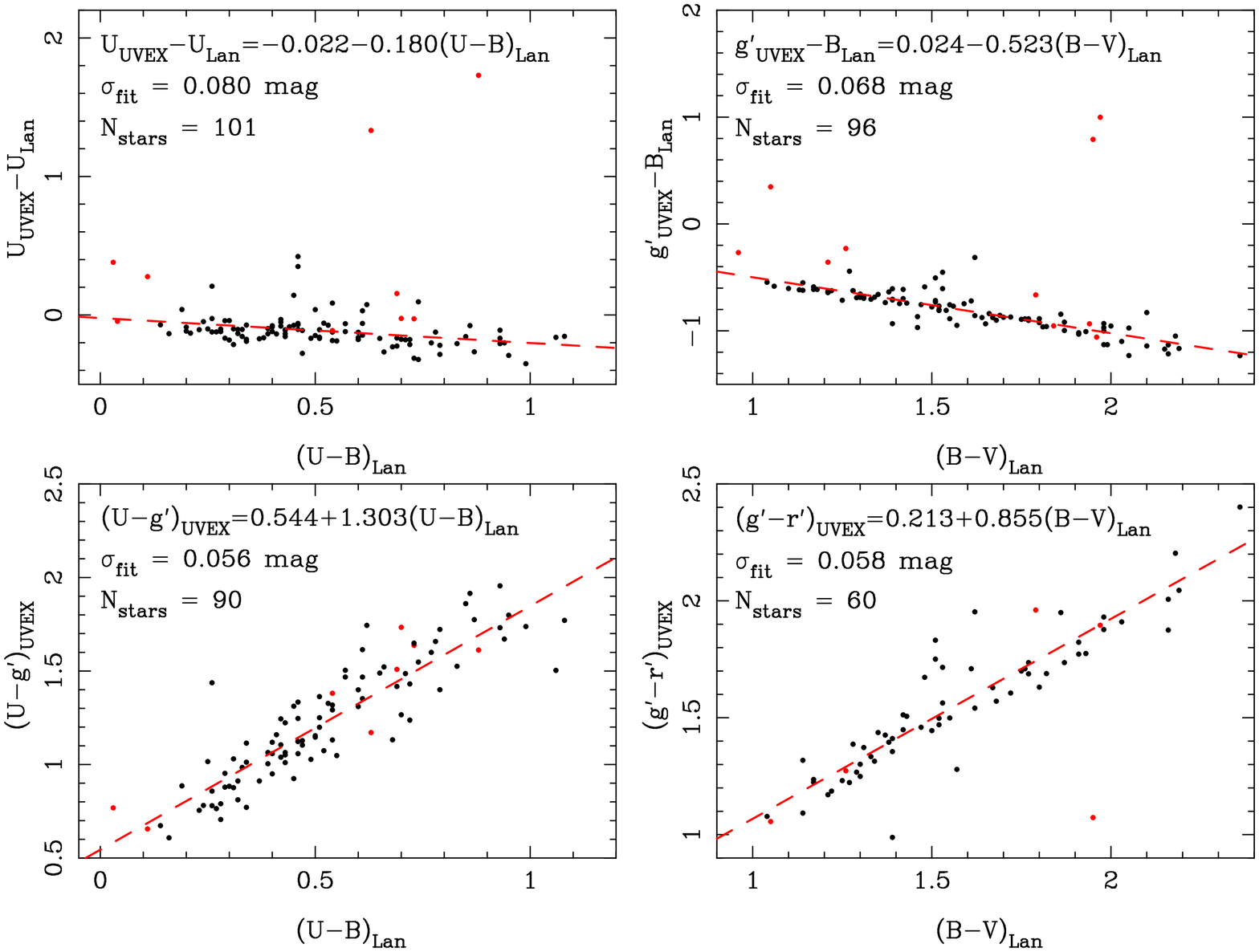}
\caption{Calibration of transformations between the \citet{mass91} $UBV$ photometry and the UVEX $U g^\prime r^\prime$ photometry. All stars are shown as black dots, with the stars listed by \citet{mass91} as `blended' shown with red dots. The best fitting transformation obtained using a least squares fit that excludes the `blended' stars are shown with red dashed lines and are listed in each panel with the number of stars used for the fit and the dispersion on the best fit.}
\label{uvex}
\end{center}
\end{figure*}

\citet{mass91} note that a number of their photometric measurements suffer from blended photometry, of which 11 are included in our census. To determine the severity of this blending and to provide alternative photometry for these sources we gathered $U g^\prime r^\prime$ photometry from UVEX, the Ultra-Violet Excess Survey \citep{groo09}, selecting sources in the magnitude range 13--19 in each filter. The $U g^\prime r^\prime$ filters are directly transferable onto the $UBV$ photometric system using transformations presented on the CASU website\footnote{Cambridge Astronomical Surveys Unit: Colours and Transformations for the Isaac Newton Telescope Wide Field Survey: http://www.ast.cam.ac.uk/$\sim$wfcsur/technical/photom/colours/}. However these transformations were derived from observations of Landolt standard fields, which are typically dominated by unreddened late-type field stars that can have different spectral energy distributions to the reddened early-type stars in our sample. We therefore calculated our own transformations by comparing the \citet{mass91} $UBV$ photometry with UVEX $U g^\prime r^\prime$ photometry and fitting new transformations (excluding the blended stars). Figure~\ref{uvex} shows the results of the least-squares straight-line fits from which we derive the transformations

\begin{equation}
U_{UVEX} - U_{Lan} = -0.22 - 0.180 (U-B)_{Lan}
\end{equation}
\begin{equation}
g^\prime - B = 0.024 - 0.523 (B-V)_{Lan}
\end{equation}
\begin{equation}
(U - g^\prime)_{UVEX} = 0.544 + 1.303 (U-B)_{Lan}
\end{equation}
\begin{equation}
g^\prime - r^\prime = 0.213 + 0.855 (B-V)_{Lan}
\end{equation}

\noindent
which are based on 101, 96, 90 and 60 stars respectively, and have dispersions of 0.080, 0.068, 0.056, and 0.058~mag respectively (no stars were found further than 3$\sigma$ from the best fits). These are in good agreement with the existing CASU transformations, differing from them by less than 10\% in the slope and typically by $\sim$0.1 in the intercepts. 

\section{The final catalog}

The full census of known OB stars in Cyg~OB2 is presented in Table~\ref{obstars} in order of increasing primary spectral sub-type and with derived observational ($A_V$), physical ($T_{eff}$, L$_{bol}$), and stellar ($M$, age) quantities \citep[the latter calculated using the][rotating evolutionary models]{ekst12}. Stars that fall to the left of the zero-aged main sequence in the HR diagram are listed as having ages of 0~Myr, while those whose positions in the HR diagram suggest ages $>$10~Myr are listed as such (see Section~\ref{s-physical} for discussion of these issues). References for spectral classifications are noted at the end of the table. The full table, which also contains $UBV$ and $JHK_s$ photometry, stellar quantities derived using the \citet{leje01} non-rotating models, along with 1$\sigma$ uncertainties on all derived properties, is available from Vizier.

\onecolumn
\newpage
\footnotesize
\begin{longtable}{cccccccccccc}
\caption{Massive stars in Cygnus OB2}\\ 
\hline 
RA & Dec & Spectral & Spectral & \multicolumn{3}{c}{Source number} & $A_V$ & log $L$ & $T_{eff}$ & $M$ & Age \\ 
\cline{5-7}
(J2000.0) & (J2000.0) & type & reference & S58 & MT91 & C02 & (mag) & ($L_\odot$) & K & ($M_\odot$) &  (Myrs)  \\ 
\hline
\endhead
\hline
\multicolumn{12}{r}{\emph{continued on next page}}\\  
\endfoot 
\endlastfoot 
20:32:03.10 & 41:15:19.9 & WC4 & vH88 &  &  & WR144 & - & 4.99 & 5.05 & $>$20 & 2-8\\ 
20:32:06.28 & 40:48:29.7 & WN7o/CE+O7V((f)) & S96,M09 &  &  & WR145 & - & 5.57 & 4.70 & $>$25 & 2-7\\ 
20:35:47.09 & 41:22:44.7 & WC6+O8III & D00,D00 &  &  & WR146 & 7.83 & 5.06 & 4.90 & $>$25 & 2-4\\ 
20:33:08.78 & 41:13:18.1 & O3If+O6V & W02,W02 & 22 & 417 &  & 6.92 & 5.82 & 4.63 & 49.9 & 4.12\\ 
20:33:14.16 & 41:20:21.5 & O3If & MT91 & 7 & 457 &  & 5.30 & 5.63 & 4.63 & 46.7 & 2.67\\ 
20:33:18.02 & 41:18:31.0 & O5III & K07 & 8C & 483 &  & 4.85 & 5.57 & 4.61 & 41.6 & 3.34\\ 
20:33:10.74 & 41:15:08.0 & O5I+O3.5III & NZ12,NZ12 & 9 & 431 &  & 6.76 & 5.85 & 4.59 & 51.6 & 4.26\\ 
20:34:08.54 & 41:36:59.3 & O5If+B0V & K12a,K12a & 11 & 734 &  & 5.40 & 5.73 & 4.59 & 43.7 & 4.56\\ 
20:33:23.46 & 41:09:12.9 & O5.5V & MT91 &  & 516 &  & 7.43 & 5.85 & 4.60 & 51.6 & 4.21\\ 
20:33:13.25 & 41:13:28.6 & O6V & MT91 &  & 448 &  & 7.22 & 5.03 & 4.59 & 28.6 & 0.76\\ 
20:31:37.50 & 41:13:21.1 & O6IV+O9III & K08,K08 & 3 &  &  & 5.84 & 5.54 & 4.59 & 38.0 & 4.18\\ 
20:33:15.18 & 41:18:50.1 & O6I+O5.5III & dB04,dB04 & 8A & 465 &  & 4.84 & 5.70 & 4.57 & 41.1 & 5.00\\ 
20:33:14.84 & 41:18:41.4 & O6.5III & MT91 & 8B & 462 &  & 5.12 & 5.48 & 4.57 & 35.2 & 4.56\\ 
20:34:44.10 & 40:51:58.0 & O6.5III & N08 &  &  & A24 & 7.29 & 5.19 & 4.57 & 29.6 & 3.41\\ 
20:33:40.88 & 41:30:18.5 & O7V & MT91 &  & 611 &  & 5.54 & 4.66 & 4.57 & 22.3 & 0\\ 
20:34:13.50 & 41:35:02.6 & O7V & MT91 & 29 & 745 &  & 5.36 & 4.94 & 4.57 & 25.4 & 1.78\\ 
20:34:29.52 & 41:31:45.5 & O7V+O9V & K08,K08 &  & 771 &  & 6.93 & 5.18 & 4.57 & 29.0 & 3.50\\ 
20:31:59.61 & 41:14:50.4 & O7V & C12 &  &  & E45 & 5.89 & 4.79 & 4.57 & 23.3 & 0.17\\ 
20:32:13.77 & 41:27:12.7 & O7III & K07 & 4 & 217 &  & 4.39 & 5.20 & 4.56 & 28.6 & 4.14\\ 
20:32:22.43 & 41:18:19.0 & O7I+O6I+O9V & B76,B76,D13 & 5 &  &  & 6.02 & 6.20 & 4.54 & 93.1 & 2.95\\ 
20:31:36.91 & 40:59:09.1 & O7I & N08 &  &  & A15 & 8.26 & 5.42 & 4.54 & 31.8 & 5.25\\ 
20:30:27.30 & 41:13:25.0 & O7I+O9I & S10,S10 &  &  & B17 & 8.08 & 5.03 & 4.54 & 24.8 & 4.17\\ 
20:33:17.49 & 41:17:09.2 & O7.5V & MT91 & 24 & 480 &  & 5.63 & 5.02 & 4.55 & 25.4 & 3.40\\ 
20:33:26.77 & 41:10:59.5 & O7.5V & MT91 &  & 534 &  & 6.36 & 4.84 & 4.55 & 23.4 & 1.77\\ 
20:32:31.50 & 41:14:08.0 & O7.5III & K12a &  & 267 &  & 8.03 & 5.51 & 4.56 & 34.7 & 5.17\\ 
20:31:10.57 & 41:31:53.0 & O8V & K08 & 1 & 059 &  & 5.18 & 5.08 & 4.54 & 25.5 & 4.46\\ 
20:32:27.67 & 41:26:21.7 & O8V & MT91 & 15 & 258 &  & 4.36 & 4.79 & 4.54 & 21.9 & 2.45\\ 
20:32:38.58 & 41:25:13.6 & O8V & H03 & 16 & 299 &  & 4.40 & 4.92 & 4.54 & 23.4 & 3.65\\ 
20:32:45.45 & 41:25:37.3 & O8V & MT91 & 6 & 317 &  & 4.54 & 5.04 & 4.54 & 24.8 & 4.28\\ 
20:32:59.17 & 41:24:25.7 & O8V & MT91 &  & 376 &  & 4.90 & 4.69 & 4.54 & 20.9 & 1.04\\ 
20:33:02.94 & 41:17:43.3 & O8V & MT91 &  & 390 &  & 6.63 & 4.93 & 4.54 & 23.5 & 3.70\\ 
20:33:13.67 & 41:13:05.7 & O8V & MT91 &  & 455 &  & 6.13 & 4.75 & 4.54 & 21.4 & 1.89\\ 
20:33:18.08 & 41:21:36.6 & O8V & MT91 &  & 485 &  & 5.32 & 4.79 & 4.54 & 21.8 & 2.40\\ 
20:33:30.43 & 41:35:57.5 & O8V & MT91 & 74 & 555 &  & 6.45 & 5.04 & 4.54 & 24.9 & 4.31\\ 
20:32:34.80 & 40:56:17.0 & O8V & N08 &  &  & A38 & 6.31 & 4.64 & 4.54 & 20.3 & 0.41\\ 
20:34:21.95 & 41:17:01.6 & O8III+O8III & K09,K09 & 73 &  &  & 6.27 & 4.62 & 4.53 & 19.6 & 1.36\\ 
20:33:02.90 & 40:47:25.0 & O8II & H03 &  &  & A20 & 8.13 & 5.58 & 4.53 & 35.0 & 5.72\\ 
20:32:50.03 & 41:23:44.6 & O8.5V & MT91 & 17 & 339 &  & 4.92 & 4.79 & 4.53 & 21.2 & 3.67\\ 
20:33:16.36 & 41:19:01.9 & O8.5V & MT91 & 8D & 473 &  & 5.13 & 4.69 & 4.53 & 20.2 & 2.59\\ 
20:33:21.04 & 41:17:40.1 & O8.5V & MT91 &  & 507 &  & 5.41 & 4.53 & 4.53 & 18.7 & 0.13\\ 
20:33:25.67 & 41:33:26.6 & O8.5V & MT91 & 25 & 531 &  & 5.46 & 4.99 & 4.53 & 23.5 & 4.88\\ 
20:31:45.39 & 41:18:26.8 & O8.5I & MT91 &  & 138 &  & 6.57 & 5.07 & 4.51 & 23.6 & 6.11\\ 
20:31:18.31 & 41:21:21.7 & O9V+B6V & K12a,K12a &  & 070 &  & 6.86 & 4.55 & 4.52 & 18.4 & 1.97\\ 
20:32:16.53 & 41:25:36.4 & O9V & MT91 & 14 & 227 &  & 4.55 & 4.66 & 4.52 & 19.1 & 3.61\\ 
20:33:09.58 & 41:13:00.6 & O9V+B9V & K09,K09 &  & 421 &  & 6.77 & 4.26 & 4.52 & 16.3 & 0\\ 
20:34:04.95 & 41:05:13.2 & O9V & MT91 & 41 & 716 &  & 6.12 & 4.45 & 4.52 & 17.5 & 0.26\\ 
20:34:09.52 & 41:34:13.4 & O9V & MT91 & 75 & 736 &  & 5.43 & 4.50 & 4.52 & 18.0 & 1.07\\ 
20:33:09.41 & 41:12:58.2 & O9V & K07 &  & 420 &  & 6.54 & 4.06 & 4.52 & 15.2 & 0\\ 
20:31:49.65 & 41:28:26.8 & O9III & K09 & 20 & 145 &  & 4.12 & 4.42 & 4.50 & 16.8 & 1.46\\ 
20:33:46.15 & 41:33:00.5 & O9I & K07 & 10 & 632 &  & 5.66 & 5.68 & 4.50 & 37.4 & 5.54\\ 
20:33:15.74 & 41:20:17.2 & O9.5V & MT91 & 23 & 470 &  & 5.13 & 4.42 & 4.50 & 16.8 & 1.35\\ 
20:33:59.57 & 41:17:36.1 & O9.5V+B0V & MT91,K12 & 27 & 696 &  & 5.81 & 4.48 & 4.50 & 17.2 & 2.58\\ 
20:30:57.70 & 41:09:57.0 & O9.5V & N08 &  &  & A26 & 6.69 & 4.69 & 4.50 & 18.7 & 5.31\\ 
20:34:16.05 & 41:02:19.6 & O9.5V & C12 &  &  & E54 & 6.23 & 4.54 & 4.50 & 17.6 & 3.62\\ 
20:31:46.00 & 41:17:27.4 & O9.5I & MT91 &  & 140 &  & 2.17 & 4.54 & 4.48 & 16.6 & 5.97\\ 
20:32:13.07 & 41:27:24.9 & B0V & MT91 &  & 213 &  & 4.30 & 4.18 & 4.49 & 14.5 & 0\\ 
20:32:59.61 & 41:15:14.6 & B0V & MT91 & 41 & 378 &  & 6.93 & 4.65 & 4.49 & 17.5 & 6.55\\ 
20:33:10.10 & 41:13:10.1 & B0V & MT91 & 51 & 425 &  & 6.55 & 4.44 & 4.49 & 16.0 & 4.32\\ 
20:33:10.34 & 41:13:06.4 & B0V & MT91 &  & 426 &  & \multicolumn{5}{c}{Unresolved}\\ 
20:33:10.50 & 41:22:22.8 & B0V+B3V & MT91,K12 &  & 429 &  & 5.45 & 3.99 & 4.49 & 13.5 & 0\\ 
20:33:37.02 & 41:16:11.4 & B0V & MT91 & 70 & 588 &  & 5.85 & 4.69 & 4.49 & 17.9 & 6.80\\ 
20:33:59.32 & 41:05:38.4 & B0V & MT91 &  & 692 &  & 5.74 & 4.14 & 4.49 & 14.2 & 0\\ 
20:30:51.12 & 41:20:21.6 & B0V & K07 &  & 020 &  & 7.20 & 4.36 & 4.49 & 15.5 & 2.65\\ 
20:32:58.79 & 41:04:29.9 & B0V+B2V & K07,K09 &  & 372 &  & 6.98 & 3.85 & 4.49 & 12.9 & 0\\ 
20:33:11.02 & 41:10:31.9 & B0V & K07 &  & 435 &  & 7.13 & 4.20 & 4.49 & 14.6 & 0\\ 
20:33:17.40 & 41:12:38.7 & B0V & K07 &  & 477 &  & 6.49 & 4.12 & 4.49 & 14.1 & 0\\ 
20:33:18.56 & 41:24:49.3 & B0V & K07 &  & 490 &  & 5.77 & 3.76 & 4.49 & 12.5 & 0\\ 
20:33:42.38 & 41:11:45.8 & B0V & K07 &  & 620 &  & 6.24 & 4.24 & 4.49 & 14.7 & 0.19\\ 
20:32:39.51 & 40:52:47.5 & B0V & C12 &  &  & E48 & 6.60 & 4.69 & 4.49 & 17.9 & 6.78\\ 
20:32:34.90 & 40:52:39.0 & B0.2V & N08 &  &  & A33 & 6.80 & 4.42 & 4.47 & 15.1 & 6.62\\ 
20:33:21.14 & 41:35:52.0 & B0III & K07 &  & 509 &  & 6.66 & 4.14 & 4.49 & 14.2 & 0\\ 
20:33:39.14 & 41:19:26.1 & B0Iab & MT91 & 19 & 601 &  & 5.72 & 5.27 & 4.49 & 26.0 & 6.74\\ 
20:34:44.70 & 40:51:46.0 & B0Ia & H03 &  &  & A27 & 7.58 & 5.63 & 4.49 & 35.2 & 5.94\\ 
20:34:58.70 & 41:36:17.0 & B0Ib+B0III & H03,K09 &  &  & A36 & 6.05 & 5.24 & 4.49 & 25.5 & 6.78\\ 
20:32:39.06 & 41:00:07.8 & B0Ia & C12 &  &  & E47 & 6.53 & 5.83 & 4.49 & 42.3 & 4.79\\ 
20:33:38.22 & 40:53:41.2 & B0Ib & C12 &  &  & E52 & 5.91 & 4.97 & 4.49 & 21.2 & 7.06\\ 
20:32:03.74 & 41:25:10.9 & B0.5V & MT91 &  & 187 &  & 5.31 & 4.05 & 4.44 & 12.3 & 3.74\\ 
20:32:27.76 & 41:28:51.9 & B0.5V & MT91 & 21 & 259 &  & 3.80 & 4.20 & 4.44 & 12.9 & 7.63\\ 
20:33:39.84 & 41:22:52.4 & B0.5V+B2.5V & K12,K12 &  & 605 &  & 4.66 & 4.10 & 4.44 & 12.5 & 5.39\\ 
20:34:06.10 & 41:08:09.6 & B0.5V+B1.5V & K12,K12 &  & 720 &  & 6.95 & 4.20 & 4.44 & 12.9 & 7.73\\ 
20:32:39.50 & 40:52:47.0 & B0.5V & N08 &  &  & A31 & 7.20 & 4.63 & 4.44 & 15.7 & $>$10\\ 
20:33:06.62 & 41:21:13.3 & B0.5V & K07 &  & 409 &  & 5.96 & 3.91 & 4.44 & 11.6 & 0\\ 
20:33:18.70 & 40:59:37.9 & B0.5III & C12 &  &  & E51 & 6.63 & 4.54 & 4.44 & 14.9 & $>$10\\ 
20:32:13.53 & 41:27:30.0 & B1V & MT91 &  & 215 &  & 4.05 & 3.58 & 4.39 & 9.2 & 0\\ 
20:32:37.03 & 41:23:05.1 & B1V & MT91 &  & 292 &  & 5.31 & 4.09 & 4.39 & 10.9 & $>$10\\ 
20:32:38.87 & 41:25:20.8 & B1V & K07 &  & 300 &  & 4.34 & 3.62 & 4.39 & 9.2 & 0.90\\ 
20:33:04.42 & 41:17:08.9 & B1V & K07 & 54 & 395 &  & 5.96 & 3.90 & 4.39 & 10.2 & $>$10\\ 
20:33:05.22 & 41:17:51.6 & B1V & K07 &  & 400 &  & 5.54 & 3.74 & 4.39 & 9.6 & 6.58\\ 
20:33:15.37 & 41:29:56.6 & B1V & MT91 &  & 467 &  & 5.41 & 3.97 & 4.39 & 10.4 & $>$10\\ 
20:33:23.24 & 41:13:41.9 & B1V & MT91 & 66 & 515 &  & 6.77 & 4.01 & 4.39 & 10.6 & $>$10\\ 
20:31:33.38 & 41:22:49.0 & B1V+B2V & K07,K12a &  & 103 &  & 6.37 & 4.04 & 4.39 & 10.7 & $>$10\\ 
20:32:14.56 & 41:22:33.7 & B1V & K07 &  & 220 &  & 5.38 & 3.61 & 4.39 & 9.1 & 0.61\\ 
20:32:21.35 & 41:18:35.5 & B1V & K07 &  & 238 &  & 5.71 & 3.45 & 4.39 & 8.7 & 0\\ 
20:32:50.69 & 41:15:02.2 & B1V & K07 &  & 343 &  & 6.55 & 4.00 & 4.39 & 10.6 & $>$10\\ 
20:32:56.66 & 41:23:41.0 & B1V & K07 &  & 365 &  & 4.81 & 3.62 & 4.39 & 9.2 & 1.14\\ 
20:33:10.46 & 41:20:57.6 & B1V & K07 &  & 428 &  & 6.00 & 3.95 & 4.39 & 10.4 & $>$10\\ 
20:33:19.16 & 41:17:44.9 & B1V & K07 &  & 492 &  & 5.67 & 3.50 & 4.39 & 8.9 & 0\\ 
20:33:23.37 & 41:20:17.2 & B1V & K07 &  & 517 &  & 5.24 & 3.79 & 4.39 & 9.8 & 9.10\\ 
20:33:42.57 & 41:14:56.9 & B1V & K07 &  & 621 &  & 6.46 & 3.79 & 4.39 & 9.8 & 9.02\\ 
20:34:04.43 & 41:08:08.4 & B1V & K07 &  & 712 &  & 6.15 & 4.18 & 4.39 & 11.4 & $>$10\\ 
20:34:24.56 & 41:26:24.7 & B1V & K07 &  & 759 &  & 5.85 & 3.68 & 4.39 & 9.4 & 3.52\\ 
20:30:39.70 & 41:08:48.0 & B0.7Ib & H03 &  &  & A23 & 7.38 & 5.42 & 4.42 & 26.3 & 7.04\\ 
20:33:47.88 & 41:20:41.7 & B1III & MT91 & 26 & 642 &  & 5.77 & 4.84 & 4.39 & 15.9 & $>$10\\ 
20:33:15.51 & 41:27:32.9 & B1III & K07 &  & 469 &  & 5.13 & 3.81 & 4.39 & 9.8 & 9.88\\ 
20:33:46.85 & 41:08:01.9 & B1III & K07 &  & 635 &  & 5.78 & 3.96 & 4.39 & 10.4 & $>$10\\ 
20:31:22.03 & 41:31:28.0 & B1I & MT91 & 2 & 083 &  & 3.98 & 4.61 & 4.39 & 14.3 & $>$10\\ 
20:33:30.81 & 41:15:22.7 & B1Ib & MT91 & 18 & 556 &  & 6.47 & 5.53 & 4.39 & 28.9 & 6.68\\ 
20:31:56.27 & 41:33:05.3 & B1.5V & MT91 &  & 169 &  & 4.41 & 3.31 & 4.36 & 7.8 & 0\\ 
20:33:05.55 & 41:43:40.1 & B1.5V & MT91 &  & 403 &  & 5.48 & 4.14 & 4.36 & 10.7 & $>$10\\ 
20:33:34.36 & 41:18:11.6 & B1.5V & MT91 &  & 575 &  & 6.73 & 4.49 & 4.36 & 12.6 & $>$10\\ 
20:33:48.88 & 41:19:40.9 & B1.5V & MT91 & 71 & 646 &  & 5.11 & 3.83 & 4.36 & 9.3 & $>$10\\ 
20:32:13.75 & 41:27:42.0 & B1.5V & K07 &  & 216 &  & 4.28 & 3.61 & 4.36 & 8.7 & $>$10\\ 
20:34:43.51 & 41:29:04.8 & B1.5III & MT91 & 30 & 793 &  & 5.73 & 4.52 & 4.36 & 12.7 & $>$10\\ 
20:32:26.50 & 41:19:13.7 & B1.5III+B1V & K07,K08 &  & 252 &  & 5.40 & 3.50 & 4.36 & 8.3 & 4.65\\ 
20:32:46.74 & 41:26:15.9 & B1.5III & K07 &  & 325 &  & 4.98 & 3.39 & 4.36 & 8.0 & 0\\ 
20:33:18.55 & 41:15:35.4 & B2Ve & K07 & 64 & 488 &  & 8.27 & 4.38 & 4.34 & 11.6 & $>$10\\ 
20:31:22.10 & 41:12:03.0 & B2V & N08 &  &  & A30 & 6.27 & 4.06 & 4.34 & 10.0 & $>$10\\ 
20:32:27.30 & 40:55:18.0 & B2V & H03 &  &  & A39 & 6.05 & 4.35 & 4.34 & 11.5 & $>$10\\ 
20:30:59.43 & 41:35:59.6 & B2V & K07 &  & 042 &  & 4.58 & 3.35 & 4.34 & 7.7 & 0\\ 
20:31:30.49 & 41:37:15.6 & B2V & K07 &  & 097 &  & 4.40 & 2.94 & 4.34 & 6.8 & 0\\ 
20:32:03.01 & 41:32:30.7 & B2Ve & K07 &  & 186 &  & 4.53 & 3.19 & 4.34 & 7.3 & 0\\ 
20:32:07.95 & 41:22:00.3 & B2V & K07 &  & 202 &  & 4.92 & 3.23 & 4.34 & 7.4 & 0\\ 
20:32:14.63 & 41:27:40.3 & B2V & K07 &  & 221 &  & 4.67 & 3.43 & 4.34 & 8.0 & 5.15\\ 
20:32:19.66 & 41:20:39.7 & B2V & K07 &  & 234 &  & 4.47 & 3.48 & 4.34 & 8.1 & 8.63\\ 
20:32:22.15 & 41:27:41.7 & B2V & K07 &  & 241 &  & 4.18 & 3.32 & 4.34 & 7.7 & 0\\ 
20:32:25.50 & 41:24:51.8 & B2V & K07 &  & 248 &  & 4.57 & 3.47 & 4.34 & 8.1 & 8.27\\ 
20:32:32.68 & 41:27:04.4 & B2V & K07 &  & 275 &  & 3.92 & 3.17 & 4.34 & 7.3 & 0\\ 
20:32:37.78 & 41:26:15.3 & B2V & K07 &  & 295 &  & 4.25 & 3.21 & 4.34 & 7.4 & 0\\ 
20:32:42.90 & 41:20:16.4 & B2V+B3V & K07,K12 &  & 311 &  & 4.95 & 3.29 & 4.34 & 7.6 & 0\\ 
20:33:22.49 & 41:22:16.9 & B2V & K07 &  & 513 &  & 5.07 & 3.33 & 4.34 & 7.7 & 0\\ 
20:33:24.78 & 41:22:04.5 & B2Ve & K07 &  & 522 &  & 4.90 & 3.34 & 4.34 & 7.7 & 0\\ 
20:33:31.68 & 41:21:46.1 & B2V & K07 &  & 561 &  & 4.67 & 3.38 & 4.34 & 7.8 & 2.13\\ 
20:33:47.63 & 41:09:06.5 & B2V & K07 &  & 639 &  & 5.94 & 3.60 & 4.34 & 8.4 & $>$10\\ 
20:33:48.83 & 41:37:39.7 & B2Ve & K07 &  & 650 &  & 5.70 & 3.35 & 4.34 & 7.7 & 0\\ 
20:31:56.90 & 41:31:48.0 & B2III & K12a &  & 174 &  & 4.39 & 3.75 & 4.34 & 8.8 & $>$10\\ 
20:32:26.10 & 41:29:39.0 & B2III & K07 &  & 250 &  & 3.91 & 3.42 & 4.34 & 8.0 & 4.66\\ 
20:32:27.26 & 41:21:56.2 & B2III & K07 &  & 255 &  & 5.06 & 3.15 & 4.34 & 7.2 & 0\\ 
20:32:30.72 & 41:07:04.1 & B2III & K07 &  & 264 &  & 3.56 & 3.37 & 4.34 & 7.7 & 1.25\\ 
20:33:11.39 & 41:17:58.9 & B2III & K07 &  & 441 &  & 5.18 & 3.31 & 4.34 & 7.6 & 0\\ 
20:33:48.40 & 41:13:14.1 & B2III & K07 &  & 645 &  & 6.19 & 3.58 & 4.34 & 8.4 & $>$10\\ 
20:30:50.75 & 41:35:06.0 & B2II & MT91 &  & 021 &  & 4.42 & 3.32 & 4.34 & 7.7 & 0\\ 
20:32:46.45 & 41:24:22.4 & B2.5V & K07 &  & 322 &  & 4.85 & 2.96 & 4.29 & 6.1 & 0\\ 
20:32:31.42 & 41:30:51.4 & B2.5V & K07 &  & 268 &  & 5.19 & 3.30 & 4.29 & 6.8 & $>$10\\ 
20:32:54.35 & 41:15:22.1 & B3V & K07 & 37 & 358 &  & 6.55 & 3.52 & 4.25 & 7.0 & $>$10\\ 
20:31:33.59 & 41:36:04.3 & B3V & K07 &  & 106 &  & 4.58 & 2.92 & 4.25 & 5.6 & $>$10\\ 
20:31:41.60 & 41:28:20.9 & B3V & K07 &  & 129 &  & 4.72 & 3.06 & 4.25 & 5.9 & $>$10\\ 
20:31:55.28 & 41:35:27.8 & B3V & K07 &  & 164 &  & 4.54 & 2.73 & 4.25 & 5.3 & 0\\ 
20:31:59.82 & 41:37:14.3 & B3V & K07 &  & 179 &  & 5.02 & 3.33 & 4.25 & 6.4 & $>$10\\ 
20:32:06.85 & 41:17:56.8 & B3V & K07 &  & 200 &  & 6.41 & 3.87 & 4.25 & 8.0 & $>$10\\ 
20:32:15.03 & 41:19:30.8 & B3V & K07 &  & 222 &  & 5.02 & 3.00 & 4.25 & 5.8 & $>$10\\ 
20:32:38.34 & 41:28:56.6 & B3V & K07 &  & 298 &  & 4.47 & 2.95 & 4.25 & 5.7 & $>$10\\ 
20:33:33.38 & 41:08:36.3 & B3V & K07 &  & 568 &  & 6.71 & 3.63 & 4.25 & 7.3 & $>$10\\ 
20:31:34.12 & 41:31:08.0 & B3IV & K07 &  & 108 &  & 4.53 & 2.64 & 4.25 & 5.1 & 0\\ 
20:32:04.74 & 41:28:44.5 & B3IV & K07 &  & 191 &  & 3.83 & 2.94 & 4.25 & 5.6 & $>$10\\ 
20:32:49.67 & 41:25:36.4 & B3III & K07 &  & 336 &  & 4.23 & 2.97 & 4.25 & 5.7 & $>$10\\ 
20:33:33.97 & 41:19:38.4 & B3I & K07 &  & 573 &  & 5.25 & 3.58 & 4.25 & 7.1 & $>$10\\ 
20:32:40.88 & 41:14:29.3 & B3.5Ia+ & MT91 & 12 & 304 &  & 10.18 & 6.09 & 4.14 & 110 & 3.0\\ 
20:32:21.76 & 41:34:24.6 & B4V & K07 &  & 239 &  & 3.92 & 2.63 & 4.22 & 4.7 & 2.66\\ 
20:32:32.34 & 41:22:57.6 & B4V & K07 &  & 271 &  & 5.14 & 3.02 & 4.22 & 5.4 & $>$10\\ 
20:33:10.27 & 41:23:44.9 & B4III & K07 &  & 427 &  & \multicolumn{5}{c}{Background}\\ 
20:31:56.23 & 41:35:12.3 & B5V & K07 &  & 170 &  & \multicolumn{5}{c}{Background}\\ 
20:32:32.54 & 41:26:46.7 & B5V & K07 &  & 273 &  & 4.53 & 2.49 & 4.20 & 4.3 & 3.92\\ 
20:33:11.81 & 41:24:05.8 & B5V & K07 &  & 444 &  & 5.25 & 3.22 & 4.20 & 5.7 & $>$10\\ 
20:33:13.37 & 41:26:39.7 & B5V & K07 &  & 453 &  & 4.27 & 2.71 & 4.20 & 4.7 & $>$10\\ 
20:33:14.34 & 41:19:33.0 & B5V & K07 &  & 459 &  & 6.65 & 3.45 & 4.20 & 6.2 & $>$10\\ 
20:33:47.58 & 41:29:57.7 & B5V & K07 &  & 641 &  & 5.71 & 3.39 & 4.20 & 6.1 & $>$10\\ 
20:33:19.26 & 41:24:44.8 & B5IV & K07 &  & 493 &  & 5.56 & 2.98 & 4.20 & 5.2 & $>$10\\ 
\hline
\label{obstars}
\end{longtable}
\noindent{\bf References:} D13 \citealt{dzib13}; C12 \citealt{come12}; D00 \citealt{doug00}; dB04 \citealt{debe04}; H03 \citealt{hans03}; K07 \citealt{kimi07}; K08 \citealt{kimi08}; K09 \citealt{kimi09}; K12 \citealt{kimi12}; K12a \citealt{kobu12}; MT91 \citealt{mass91}; N08 \citealt{negu08}; NZ12 \citealt{naze12}; S96 \citealt{smit96}; S10 \citealt{stro10}; vH88 \citealt{vand88}; W02 \citealt{walb02}; B76 \citealt{boha76}; M09 \citealt{munt09}.\\

\end{appendix}

\end{document}